\documentclass[a4paper]{article}
\usepackage[utf8]{inputenc}
\usepackage{biblatex}
\usepackage{sectsty}
\usepackage{graphicx}
\usepackage{hyperref}
\usepackage{siunitx}
\usepackage[acronym]{glossaries}
\usepackage{ragged2e}
\usepackage{changepage}
\usepackage{authblk}
\usepackage{geometry}
\usepackage[dvipsnames]{xcolor}
\usepackage{physics}
\usepackage{amssymb} %\Gamma is not defined
\usepackage{enumitem}
\usepackage{csquotes}
\usepackage{multicol}

\graphicspath{
  {./}
  {./sessions files/Mental-health/}
}

\newacronym{gw}{GW}{gravitational wave}
\newacronym{esa}{ESA}{European Space Agency}
\newacronym{lisa}{LISA}{Laser Interferometer Space Antenna}
\newacronym{lpf}{LPF}{LISA Pathfinder}
\newacronym{tdi}{TDI}{time-delay interferometry}
\newacronym[longplural={spacecraft},shortplural={SC}]{sc}{SC}{spacecraft}
\newacronym{tm}{TM}{test mass}
\newacronym[longplural={movable optical sub-assemblies}]{mosa}{MOSA}{movable optical sub-assembly}
\newacronym[longplural={supermassive black hole binaries}]{smbhb}{SMBHB}{supermassive black hole binary}
\newacronym[longplural={white dwarf binaries}]{wdb}{WDB}{white dwarf binary}
\newacronym{emri}{EMRI}{extreme mass ratio inspiral}
\newacronym{smbh}{SMBH}{supermassive black hole}
\newacronym{em}{EM}{electromagnetic}
\newacronym{sgwb}{SGWB}{stochastic gravitational waves background}
\newacronym{bbn}{BBN}{Big Bang nucleosynthesis}
\newacronym{pbh}{PBH}{primordial black hole}
\newacronym{dfacs}{DFACS}{drag-free attitude control system}
\newacronym{lca}{LCA}{LISA core assembly}
\newacronym{grs}{GRS}{gravitational reference sensor}
\newacronym{ifo}{IFO}{interferometer}
\newacronym{snr}{SNR}{signal-to-noise ratio}
\newacronym{psd}{PSD}{power spectral density}
\newacronym{ldc}{LDC}{LISA data challenge}
\newacronym[longplural={double white dwarves}]{dwd}{DWD}{double white dwarf}
\newacronym[longplural={black-hole binaries}]{bhb}{BHB}{black-hole binary}

\sectionfont{\centering}
\title{Legacy of the First Workshop on Gravitational Wave Astrophysics for Early Career Scientists}

\author[1]{Jean-Baptiste Bayle}
\affil[1]{\small Jet Propulsion Laboratory, California Institute of Technology, Pasadena, CA 91125, United States}

\author[2]{B\'eatrice Bonga}
\affil[2]{\small Institute for Mathematics, Astrophysics and Particle Physics, Radboud University, 6525 AJ Nijmegen, The Netherlands}

\author[3,4]{Daniela Doneva}
\affil[3]{\small Theoretical Astrophysics, Eberhard Karls University of T\"ubingen,  T\"ubingen 72076, Germany}
\affil[4]{\small INRNE - Bulgarian Academy of Sciences, 1784 Sofia, Bulgaria}

\author[5]{Tanja Hinderer}
\affil[5]{\small Institute for Theoretical Physics, Utrecht University, Princetonplein 5, 3584 CC Utrecht, The Netherlands}

\author[6]{Archisman Ghosh}
\affil[6]{\small Ghent University, Proeftuinstraat 86, 9000 Gent, Belgium}

\author[7]{Nikolaos Karnesis}
\affil[7]{\small Department of Physics, Aristotle University of Thessaloniki, Thessaloniki 54124, Greece}

\author[8]{Mikhail Korobko}
\affil[8]{\small Institut f\"ur Laserphysik \& Zentrum f\"ur Optische Quantentechnologien, Universit\"at Hamburg,
Luruper Chaussee 149, 22761 Hamburg, Germany}

\author[9]{Valeriya Korol}
\affil[9]{\small Institute for Gravitational Wave Astronomy \& School of Physics and Astronomy, University of Birmingham, Birmingham, B15 2TT, UK}

\author[10,11]{Elisa Maggio}
\affil[10]{\small Dipartimento di Fisica, ``Sapienza'' Universit\`a di Roma \& Sezione INFN Roma1, Piazzale
Aldo Moro 5, 00185, Roma, Italy}
\affil[11]{\small Max-Planck-Institut f\"ur Gravitationsphysik, Albert-Einstein-Institut, 
Am M\"uhlenberg 1, 14476 Potsdam-Golm, Germany}

\author[12]{Martina Muratore}
\affil[12]{\small Dipartimento di Fisica, Università di Trento and Trento Institute for Fundamental Physics and Application/INFN, 38123 Povo, Trento, Italy}

\author[13]{Arianna I. Renzini}
\affil[13]{\small LIGO  Laboratory, and Cahill astronomy department,  California  Institute  of  Technology,  Pasadena,  CA  91125,  USA}

\author[14]{Angelo Ricciardone}
\affil[14]{\small Dipartimento di Fisica e Astronomia ``G. Galilei",
Universit\`a degli Studi di Padova, via Marzolo 8, I-35131 Padova, Italy \& INFN, Sezione di Padova,
via Marzolo 8, I-35131 Padova, Italy}

\author[15]{Sweta Shah}
\affil[15]{\small Max-Planck-Institut f\"ur Gravitationsphysik (Albert-Einstein-Institut), D-30167 Hannover, Germany}

\author[16]{Golam Shaifullah}
\affil[16]{\small Dipartimento di Fisica ``G. Occhialini'', Università di Milano-Bicocca, Piazza della Scienza 3, 20126, Milano, Italy}

\author[17,18]{Lijing Shao}
\affil[17]{\small Kavli Institute for Astronomy and Astrophysics, Peking University, Beijing 100871, China}
\affil[18]{National Astronomical Observatories, Chinese Academy of Sciences, Beijing 100012, China}

\author[11]{Lorenzo Speri}
\author[19]{Nicola Tamanini\footnote{General coordinator and main contact (nicola.tamanini@l2it.in2p3.fr).}}
\affil[19]{\small Laboratoire des 2 Infinis - Toulouse (L2IT-IN2P3), Universit\'e de Toulouse, CNRS, UPS, F-31062 Toulouse Cedex 9, France}

\author[20]{David Weir}
\affil[20]{\small Department of Physics and Helsinki Institute of
  Physics, P.O. Box 64, 00014 University of Helsinki, Finland}

\date{}

\begin{document}

\maketitle
\thispagestyle{empty}

\newpage

\begin{abstract}
Gravitational wave science is a dynamical, fast-expanding research field founded on results, tools and methodologies drawn from different research areas and communities.
Early career scientists entering this field must
learn and combine knowledge and techniques from
a range of disciplines.
The Workshop on \textit{Gravitational--Wave Astrophysics for Early Career Scientists} (GWÆCS), held virtually in May 2021, planted the
seeds
of an interdisciplinary, well-connected and all-inclusive
community of early career scientists working on gravitational waves,
able
to exchange relevant information and ideas, build a healthy
professional and international environment, share and learn valuable
skills, and ensure that
ongoing
research
efforts are perpetuated and expanded in order to attain the main scientific goals envisioned by the whole community.
GWÆCS was the first event unifying early career scientists belonging
to different communities, historically associated
with
different large-scale gravitational wave experiments.
It provided a
broad perspective on the future of gravitational waves,
offered training on soft and transferable skills and
allowed ample time for informal discussions between early career scientists and well-known research experts.
The essence of those activities is summarised and collected in
the present document, which presents a recap of each session of the
workshop and aims
to provide all early career scientists with a long-lasting, useful reference which constitutes the legacy of all the ideas that circulated at GWÆCS.
\end{abstract}

\newpage
\begin{figure}
    \centering
    \includegraphics[width=\textwidth]{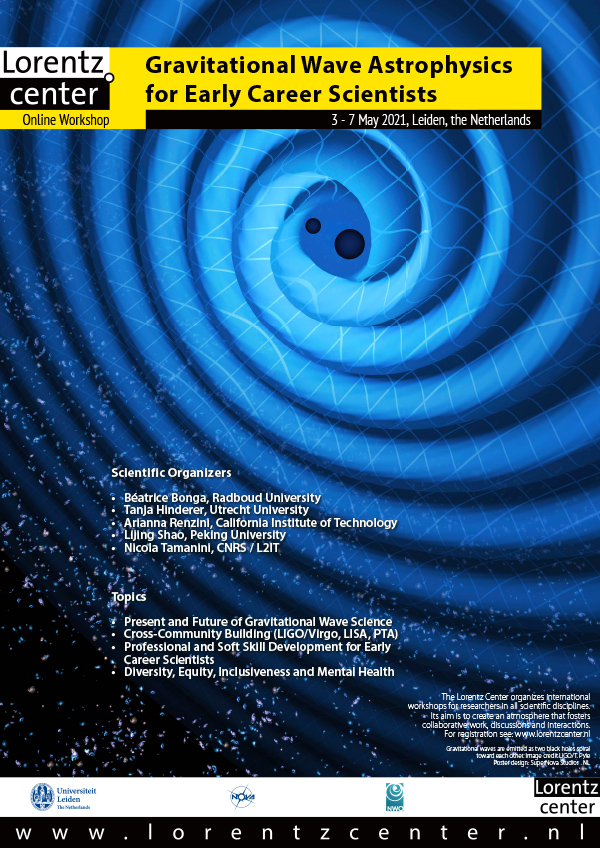}
    %\caption{Caption}
    %\label{fig:poster}
\end{figure}

\newpage
\tableofcontents

\newpage
\section*{Preface}
\addcontentsline{toc}{section}{Preface}
The discovery of gravitational waves (GWs) in 2015 changed our way of exploring the Universe, and inaugurated a new research field which is growing at an increasing pace worldwide.
The promise of important discoveries in astrophysics, cosmology and
fundamental physics attracts
students, researchers, and entire university departments from various backgrounds to this newly established field.
Early-career scientists (ECSs) working in the field of GWs today must thus build their research curricula by learning and combining knowledge, tools and methodologies from different disciplines.
They represent the future generation that will have the responsibility to complete and improve the current large-scale GW experiments, and to push forward the related theoretical and data analysis efforts needed to maximise their science return and drive future endeavours.
ECSs therefore are in a unique position to: contribute to the current decisions on the road map for the field; shape data and publication policies; build a diverse and inclusive community; and preempt important issues inherent to ECSs themselves and to the future of the field.

The current large-scale, international GW experiments are clustered around three main areas, based on different GW detection approaches: ground-based detectors, space-borne detectors and pulsar timing arrays.
While each of these clusters has a well-connected internal community, the networking between them and the connections to the wider theoretical and data analysis GW community can definitely be improved.
In particular, to fully take advantage of all future opportunities that GWs will offer, it is imperative to expose ECSs to knowledge and methods usually employed outside their main sphere of research.
We are still at the beginning of the GW astronomy era, and it will become crucial for the success and cohesion of the field that the community becomes better connected and better aware of its own possibilities.
New detectors will come online in the coming decades, and it will be impossible to carry out these large-scale observations and maximize their scientific impact without close collaboration between different fields of research and communities.
Since the ECSs of today will be the leaders of tomorrow's endeavours, now is the right time to start building the GW community of the future.

These were the premises that underlay the scope and rationale of the workshop on \textit{Gravitational-Wave Astrophysics for Early Career Scientists} (GW{\AE}CS), which was held virtually on 3-7 May 2021 with the help of the \textit{Lorentz Center}.
GW{\AE}CS was designed to build networks across different communities and advance the professional development of ECSs working on GWs.
The principal goal of GW{\AE}CS was to plant the seeds to grow a community of ECSs with the scope of
\begin{enumerate}
    \item  Becoming a strong and constructive voice for the future GW community,
    \item Addressing the needs of ECSs and foster a healthy professional environment,
    \item Sharing career opportunities and learn valuable professional soft skills,
    \item Ensuring that the international and interdisciplinary GW community shares a common scientific language, builds long-lasting communication channels and exchanges relevant research knowledge and tools for the benefit of its own future.
\end{enumerate}
An extract from the workshop program submitted to the Lorentz Center clarifies the practical objectives of GW{\AE}CS:
\begin{quote}
    \textit{The GW{\AE}CS workshop will be considered a success if the participants acquire a broader perspective on the common challenges and questions in the field, establish new networks with other ECSs and senior members, acquire new skills and useful knowledge for their professional and personal development, and share the resources they will develop as deliverables of the workshop with other ECSs. The participants of this workshop will be the first ambassadors of a constructive and all-inclusive community of scientists determined to explore and discover the universe with GWs.}
\end{quote}

GW{\AE}CS offered a wide perspective on the future of GWs, focusing in particular on the expected scientific gain from forthcoming large-scale experiments and on the common effort needed to successfully complete them.
Presentations from well-known experts from the different GW communities world-wide were followed by dedicated discussions aimed at advancing the knowledge and gauging the expectations of ECSs regarding the future of GWs.
Equally important, complementing the development of scientific topics, GW{\AE}CS offered training on soft and transferable skills.
In particular, it targeted leadership, outreach activity design, education on diversity and inclusion, awareness about well-being and mental-health issues, how to search and interview for academic and industry positions, and the different responsibilities of a faculty member.
\begin{figure}
    \centering
    \includegraphics[width=\textwidth]{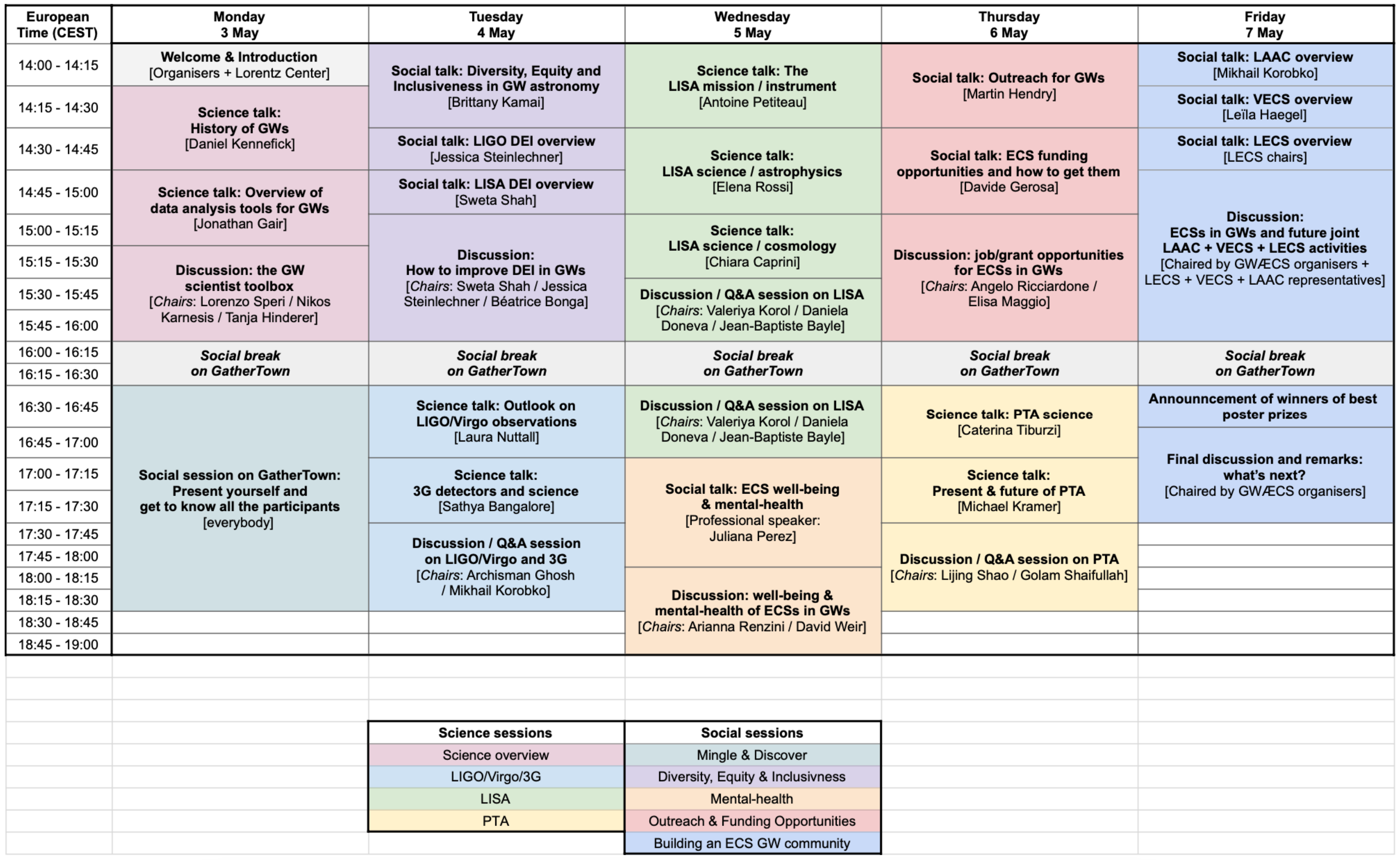}
    \caption{Program of the GW{\AE}CS online workshop}
    \label{fig:gwaecs_progr}
\end{figure}
The workshop program represented a mixture of overview talks on specific scientific and social topics, with an extended time for Q\&A sessions, discussions and training (see Fig.~\ref{fig:gwaecs_progr}).
50\% of the time was dedicated to science and 50\% to social issues and professional training relevant for ECSs.
The majority of the scheduled time was reserved for discussions, while the scope of the selected talks consisted in providing basic information, suggesting useful inputs and sparking general curiosity for the subsequent debates. 
The overview talks helped to connect the participants from different sub-fields, and to emphasize overarching challenges and visions for the GW community.
The science and social talks were given either by senior invited academics or by professional speakers, while the discussions were all chaired by experienced ECSs (junior faculty or senior postdocs) in an open and inclusive way.
Moreover, discussion sessions were organised to collect different visions for the future of GWs with the scope of producing a constructive dialogue aimed at providing a smooth and shared roadmap for the evolution of the field.

GW{\AE}CS had two main deliverables: a community one and a material one.
The first intangible deliverable was to establish the first step for a wide network of ECSs able to organise themselves outside big international collaborations clustered around large-scale GW experiments.
The first objective is then to build an organisation connecting all ECSs working in GWs in order to share relevant information, discuss issues not necessarily inherent to a specific GW experiment or topic, and foster long-lasting relationships to prepare a united future GW community.
This effort is currently underway and in September 2021 a first online meeting among all ECSs which actively took part in the activities of GW{\AE}CS, together with representatives from the ECSs of different GW communities\footnote{In particular representatives from the LISA Early Career Scientists (LECS) group, the Virgo Early Career Scientists (VECS) group, the LIGO Academy Advisory Council (LAAC) and the Pulsar Timing Array (PTA) community.}, was organised.
As a result of this meeting, a collective emerged towards the creation of an overall structure connecting ECSs in all different GW communities with the concrete scope of organising and coordinating activities aimed at fulfilling the objective outlined above, in perfect alignment with the spirit of GW{\AE}CS.

On the other hand, the material deliverable of GW{\AE}CS was the writing of a document summarising all the most important discussions that happened during the workshop.
The present document constitutes the completion of this objective.
It represents a collection of short summaries each presenting a session that took place during the workshop.
Each one of these summaries contains not only a recap of the overview talks that were given by well-known experts, but also, and more crucially, an extract from the interesting discussion between ECSs and speakers that took place right after the overview talks.
These discussions were the main focus and strength of GW{\AE}CS as they gave voice to ECSs and allowed them to ask everything they wanted to know on the specific topic under consideration.
The idea of the present document is to collect in a single place these questions, comments and impressions for the benefit of the whole GW community, and specifically of its ECSs.

The summary of each GW{\AE}CS session is written by the session chairs with the help of other ECSs, and is based on the abundant material collected during the workshop, including recordings of the sessions, extracts from the live chat and messages exchanged off-line.
Each summary has been reviewed first by the invited speakers who contributed to give a talk in the corresponding session, and then by all the participants of GW{\AE}CS.
The overall coordination of the effort needed to complete all the summaries and collect them into the present document has been provided by the GW{\AE}CS organisers, led by Nicola Tamanini.
Each one of the summaries well represents the mood and characteristics of its own session and for this reason each of them presents a unique style arising from a combination of the writers' approach, the treated topics and presentations given in each session.
For this reason this document can come across as a collection of separated and disconnected reports, however one must keep in mind the overall objective that connects all of them together and links them to the spirit of GW{\AE}CS: namely, the idea of starting to build a community that in time will be able not only to address the most important questions in GW science, but also to do so by respecting and possibly improving the lives of the very people involved in the process.

\subsection*{GW{\AE}CS workshop feedback and selected comments}

\begin{figure}
    \centering
    \includegraphics[width=0.8\textwidth]{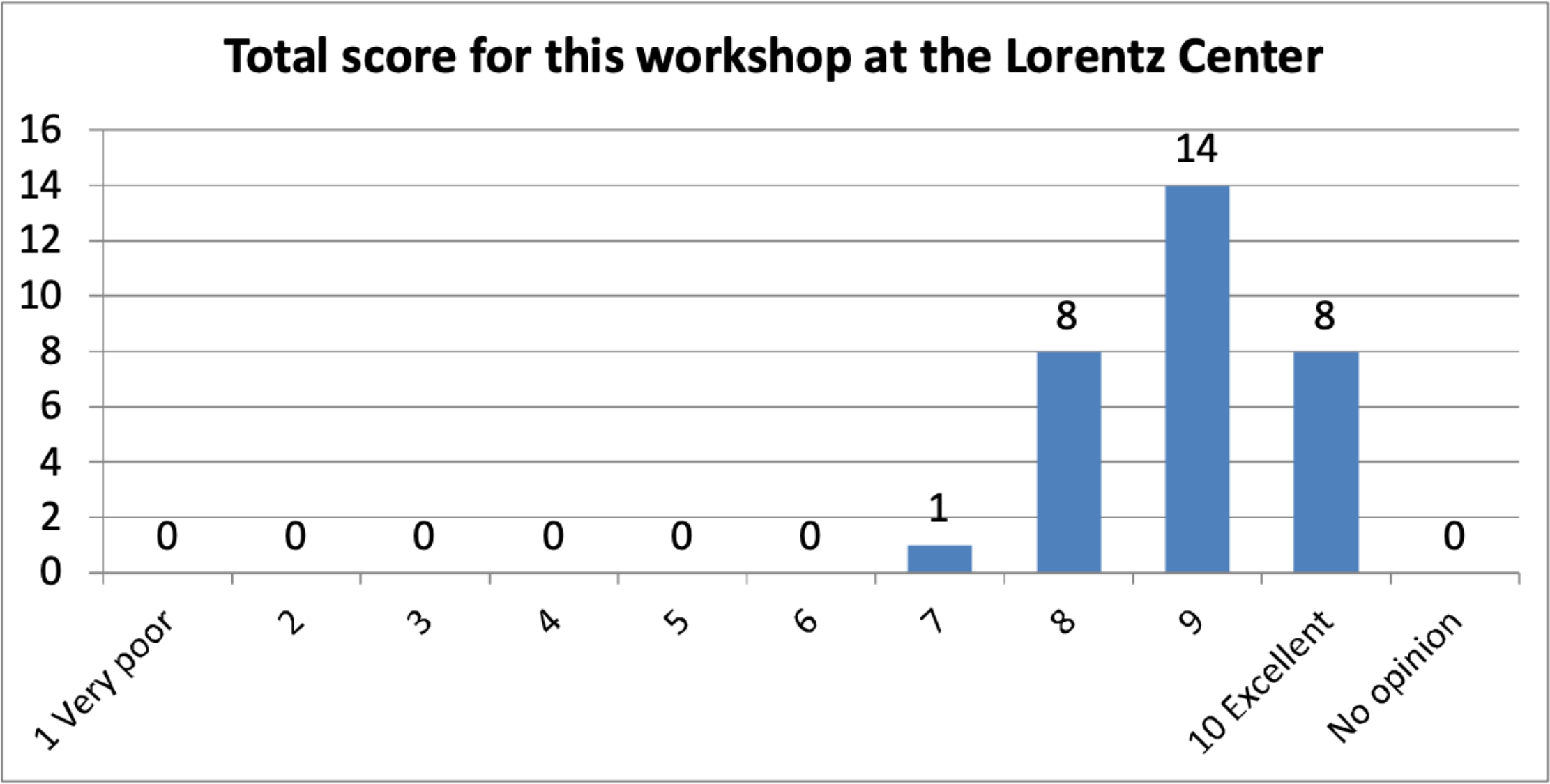}
    \caption{Overall feedback from the participants of the GW{\AE}CS online workshop.}
    \label{fig:feedback}
\end{figure}

The Lorentz Center prepared and submitted a questionnaire to all participants to collect general feedback about their experience at GW{\AE}CS.
As shown by Fig.~\ref{fig:feedback}, which reports the overall score that respondents gave to the workshop, GWÆCS was a great success.
It attracted more than 250 registered participants of which up to 80 had the opportunity to follow talks live and participate in the discussions and Q\&A session, while the rest of the participants had access to the recordings of each session and to asynchronous communication services (in particular a Slack platform dedicated to the workshop).
The comments here below reflects some of the written feedback that has been collected through the questionnaire prepared by the Lorentz Center.\\

\begin{adjustwidth}{0.2\textwidth}{0pt}
\begin{flushright}
\textit{What I liked the most about this workshop was the soft skills sections, because it was brand new for me to have this on an event with discussions about these topics. Nevertheless I loved the fact that the talks were meant to be an overview to motivate discussions. As we had a very large time for discussions we could expose our ideas and doubts in a more extensive way than in a normal 2-5 min Q\&A after a lectures in a large conference. The structure of the workshop was (almost) perfect for me. If it is not a limitation for the use of the place I would suggest the same format but 2 weeks for inclusions of more topics.}
\end{flushright}
\end{adjustwidth}
\medskip

\begin{adjustwidth}{0pt}{0.2\textwidth}
\begin{flushleft}
\textit{This was a workshop designed partly to and for a network of early career scientists in the field. In that respect it was very successful. It also did an extremely good job of discussing "soft skills'' which do not usually come up in most scientific meetings.}
\end{flushleft}
\end{adjustwidth}
\medskip

\begin{adjustwidth}{0.2\textwidth}{0pt}
\begin{flushright}
\textit{Honestly one of the best workshops I've attended online!}
\end{flushright}
\end{adjustwidth}
\medskip

\begin{adjustwidth}{0pt}{0.2\textwidth}
\begin{flushleft}
\textit{The larger focus on the discussion section made the online experience a lot better.}
\end{flushleft}
\end{adjustwidth}
\medskip

\begin{adjustwidth}{0.2\textwidth}{0pt}
\begin{flushright}
\textit{The discussion sessions were very interesting and useful for early career scientists. I like the fact that the talks dealt with several aspects related the gravitational-wave science that are not addressed usually in other conferences, like the mental health and the cooperation between different groups.}
\end{flushright}
\end{adjustwidth}
\medskip

\begin{adjustwidth}{0pt}{0.2\textwidth}
\begin{flushleft}
\textit{One of the best workshops so far in my early carrier!}
\end{flushleft}
\end{adjustwidth}
\medskip

\begin{adjustwidth}{0.2\textwidth}{0pt}
\begin{flushright}
\textit{Well organised. Really enjoyed the soft skills sessions - those were most valuable to me.}
\end{flushright}
\end{adjustwidth}
\medskip

\begin{adjustwidth}{0pt}{0.2\textwidth}
\begin{flushleft}
\textit{This is by far one of the best workshops I've been to online. I am not an early career researcher, so the workshop was not aimed at me as such. However I enjoyed the talks, they were at a great introductory level and I really liked the emphasis on ECSs.}
\end{flushleft}
\end{adjustwidth}
\medskip

\newpage
\section*{Acknowledgements \& list of endorsers}
\addcontentsline{toc}{section}{Acknowledgements \& list of endorsers}

GWÆCS would have never been possible without the opportunity provided by the Lorentz Center and the help of its professional and friendly staff.
In particular a huge thanks goes to Daniëlle van Rijk (the workshop coordinator), and to Jacqueline de Boer, Henriette Jensenius, Jantien Schuijer and Wendy van der Linden whose assistance with organisational and administrative issues was fundamental to the success of GWÆCS.
The workshop was financed by the Lorentz Center, by the \textit{Nederlandse Onderzoekschool Voor Astronomie} (NOVA) and by the COST Action CA16104 ``Gravitational waves, black holes and fundamental physics'' (GWverse), supported by COST (European Cooperation in Science and Technology).
The GWÆCS organisers are extremely grateful to the senior advisory board composed of Bangalore Sathyaprakash, Elena Maria Rossi and Michael Kramer for providing useful inputs during the organisation of the GWÆCS workshop and during the workshop itself.
We would also like to thank Monica Colpi and Stoytcho Yazadjiev for providing useful comments on parts of this document.

\subsection*{List of endorsers}

All the participants of GWÆCS were asked to endorse this document and provide feedback on it. Here below we report the list of all participants who decided to endorse this work.

\medskip

\begin{multicols}{2}
\noindent
Alberto Mangiagli\\
Alexander Saffer\\
Ali Seraj\\
Ali Yoonesyaan\\
Alireza Allahyari\\
Andrea Chincarini\\
Andrei Ieronim Constantinescu\\
Andrew R.~Williamson\\
Anslyn John\\
Chiara Caprini\\
Daniel Hartwig\\
Davide Gerosa\\
Dhruba Jyoti Gogoi\\
Elena Maria Rossi\\
Fei Xu\\
Garvin Yim\\
Geoffrey Mo\\
Gopi Kamlesh Patel\\
Gregorio Carullo\\
Hossein Salahshoor Gavalan\\
Jam Sadiq\\
Jose María Ezquiaga\\
Julio César Martins\\
Jyatsnasree Bora\\
Laura K. Nuttall\\
Laura Sberna\\
Leïla Haegel\\
Luis Felipe Longo Micchi\\
Maria Elidaiana da Silva Pereira\\
Michele Mancarella\\
Paolo Cremonese\\
Philip Lynch\\
Rajesh Kumar Dubey\\
Răzvan Balașov\\
Shobhit Ranjan\\
Surabhi Sachdev\\
T.~S.~Sachin Venkatesh\\
Thomas Kupfer\\
Vojtěch Witzany
\end{multicols} 

\newpage
\section*{List of invited speakers at GWÆCS}
\addcontentsline{toc}{section}{List of invited speakers at GWÆCS}
The great success of GWÆCS could have never been happened without the insightful input conveyed by all the wonderful talks delivered by the invited speakers.
Their active participation in the highly dynamical discussions and Q\&A sessions, together with their convivial and informal approach, produced the perfect environment for ECSs to be well at ease to ask any kind of questions and created in this way deeply instructive moments for all participants.
The invited speakers also reviewed and provided feedback on the summaries written by the session chairs, which clearly helped improving the overall quality of these records.
For all these services to GWÆCS we thank them immensely and list their names here below.

\begin{itemize}
    \item \textbf{Chiara Caprini}\\ APC - AstroParticule et Cosmologie, Université de Paris
    \item \textbf{Jonathan Gair}\\ Albert Einstein Institute, Potsdam
    \item \textbf{Davide Gerosa}\\ Universit\`a degli Studi di Milano-Bicocca \& INFN
    \item \textbf{Martin Hendry}\\ University of Glasgow
    \item \textbf{Britanny Kamai}\\ University of California Santa Cruz and Caltech
    \item \textbf{Daniel Kennefick}\\ University of Arkansas
    \item \textbf{Michael Kramer}\\ Max-Planck-Institut für Radioastronomie \& University of Manchester
    \item \textbf{Laura Nuttall}\\ University of Portsmouth
    \item \textbf{Juliana Perez}\\ California Institute of Technology, Pasadena, California
    \item \textbf{Antoine Petiteau}\\ APC - AstroParticule et Cosmologie, Université de Paris
    \item \textbf{Elena Maria Rossi}\\ Leiden Observatory
    \item \textbf{Bangalore Sathyaprakash}\\ The Pennsylvania State University
    \item \textbf{Sweta Shah}\\ Albert Einstein Institute, Hannover
    \item \textbf{Jessica Steinlechner}\\ Maastricht University
    \item \textbf{Caterina Tiburzi}\\ ASTRON - the Netherlands Institute for Radio Astronomy
\end{itemize} 

\newpage
\part{Social Sessions}

\newpage
\begin{refsection}
\setcounter{figure}{0}
\setcounter{footnote}{0} 
\section{Diversity, Equity and Inclusiveness}
\begin{center}
    \textbf{Session chairs \& writers:}\\
    B\'eatrice Bonga\footnote{Institute for Mathematics, Astrophysics and Particle Physics, Radboud University, 6525 AJ Nijmegen, The Netherlands}
\end{center}

\begin{center}
    \textbf{Invited speakers:}\\
    Britanny Kamai\footnote{Department of Astronomy and Astrophysics, University of California Santa Cruz, Santa Cruz, California 95064, USA}$^{\rm ,}$\footnote{Department of Mechanical and Civil Engineering, California Institute of Technology, Pasadena, California 91125, USA}\\
    Jessica Steinlechner\footnote{Maastricht University, P.O. Box 616, 6200 MD Maastricht, Netherlands}\\
    Sweta Shah\footnote{Max-Planck-Institut für Gravitationsphysik (Albert-Einstein-Institut), D-30167 Hannover, Germany}
\end{center}

Early-career scientists feel increasingly connected to societal issues such as climate change, sexism and racism. The participants of GWÆCS are no different. A natural arena for us to tackle such issues is our own scientific community. Therefore, we had a dedicated session on the important topic of Diversity, Equity and Inclusion (DEI) during GWÆCS. This session was very well attended and a lively discussion followed the presentations on this topic. \\

This session started off with an overview by Dr.~Britanny Kamai on DEI, its importance, and how to be an ally.  Kamai is an astrophysicist working on the intersection of cosmology, gravitational wave detectors and seismic metamaterials. She also has become a DEI advocate. In that capacity, she served on domestic and international advisory boards at organizations such as NASA, ESA, APS, AAAS and SACNAS to help shape a healthier future for gravitational wave astrophysics, bridge programs, science policy, and STEM. Her presentation was followed by a short overview on the different DEI initiatives within the LVC and LISA by Dr.~Jessica Steinlechner and Dr.~Sweta Shah. The session ended with a Q\&A with all the speakers.

\subsection*{What is DEI?}
DEI stands for diversity, equity and inclusion. Diversity simply refers to the presence of differences within a given setting. Equity is the process of ensuring that processes and programs are impartial, fair and provide equal possible outcomes for every individual. Inclusion is the practice of ensuring that people feel a sense of belonging. DEI is not about aesthetics, but about the health of our scientific community. Often it is presented as important, yet something that receives little attention and can be thought of after working hours. However, it is our science. 

There are many axes of identity and consequently of diversity. Examples are gender, ethnic and racial identity, LGBTQIA+, (dis)-abilities and religion. While in our own scientific community, there is a gender gap, which deserves all our attention, our focus on diversity should not be limited to this axis of identity. In particular, diversity is not a community of men and able-bodied, white women from privileged socio-economic backgrounds.

DEI is about creating the optimal scientific working environment for everyone so that every individual feels supported and valued for their intellectual contributions. In such an environment, you have access to participate, be listened to and are invited into spaces to create and innovate thanks to/despite your identity. These environments are diverse and everyone’s cultural identity and gender expression is supported. In contrast, in a disastrous working environment, some feel erased and devalued for their contributions, are mis-gendered, and experience forms of racism and/or sexism. There is no way for all members to actively or passively contribute to the research. If these issues are raised, victims are being blamed and invalidated, no action is taken around what happened and the injured individuals continue to be intentionally or unintentionally left out. You could be in the same room as one of your fellow scientists and experience a warm and welcoming scientific environment because your identities are supported and your voice is valued, while for others the environment could be an apocalyptic wasteland. We need to learn to listen and treat others how they want to be treated. 

Lack of diversity in science is a symptom of a much deeper societal problem: systemic oppression, i.e. the prolonged and unjust treatment of marginalized groups within society. Systemic oppression exists at the level of institutions in the form of harmful policies and practices and across structures such as education, health, transportation and economy. Fortunately, systemic oppression can be undone through recognition of inequitable patterns and intentional action to interrupt inequity and create more democratic processes and systems. In other words, systemic oppression can be undone through systemic liberation (see Fig. 1). The goal is to create conditions without barriers so that people can be themselves and fully participate. 

\begin{figure}
    \centering
    \includegraphics[width=\textwidth]{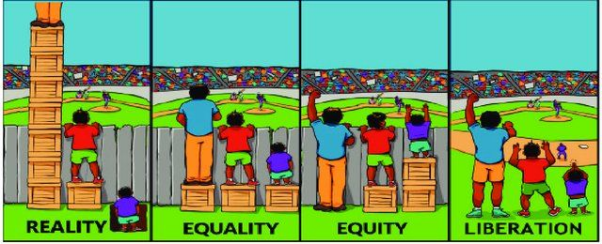}
    \caption{The difference between equality, equity and liberation. [Figure taken from “The A-Z of Social Justice Physical Education: Part 1”, Journal of Physical Education Recreation \& Dance 91(4)]}
    \label{fig:Equality_vs_liberation}
\end{figure}

To achieve scientific liberation, we need to update our value system in the gravitational wave community. Supporting diverse populations is not charity, second-rate, for fun or to do in your spare time: it is science. Research shows that the most innovative science comes from the most diverse groups of people. Therefore, as early-career scientists, we value top-level high-quality innovative science AND uplifting, compassionate, supportive and diverse working environments.  

\subsection*{Action items for us all
}
We can all contribute to a better work environment for our colleagues and colleagues-to-be. We achieve this by making DEI part of our everyday life and not simply attending a one-off talk about DEI. Here is a list of concrete action items for you to consider (in no particular order):
\begin{enumerate}
    \item Inform yourself about issues regarding DEI;
    \item Be sincere and talk to others about these issues;
    \item Make an effort to pronounce the names of the people you encounter;
    \item Speak up when you see someone being marginalized and be an ally (if you feel insecure about this, consider attending bystander intervention training);
    \item Make DEI part of your weekly research group meetings (e.g. discuss scientific literature on this topic, discuss relevant recent events in the news or possible action items members can take);
    \item Become a member of a DEI committee in your institution or the scientific collaboration you are part of;
    \item Speak up on social media about DEI;
    \item Do outreach to diverse populations;
    \item Create communities of support to ensure no one feels alone in their journey and share tips on how you have navigated through;
    \item When suggesting or inviting speakers for seminars and conferences, be aware of a diverse list;
    \item When receiving an invitation for a panel or conference that is not diverse, consider declining the invitation and explain why;
    \item Consider joining a March that you support.
\end{enumerate}

\subsection*{Resources across GW scientific collaborations}

\subsubsection*{DEI in LSC}
Within the LSC, there are several places with responsibilities related to DEI:
\begin{itemize}
    \item The \href{wiki.ligo.org/LSC/Diversity)}{DEI committee} \cite{DEI-LSC}, composed of appointed members, coordinates the projects and wider efforts undertaken by the \href{wiki.ligo.org/LSC/Diversity/DiversityGroup}{LSC diversity working group} \cite{Diversity-LSC}, membership of which is open to the entire LSC. The DEI Committee originated from the revised LSC By-laws (2020). While the DEI committee is relatively new, the diversity working group has been existing for almost ten years. The DEI committee works closely with the Speakers Committee and with the LAAC.
    \item The \href{laac.docs.ligo.org}{LAAC} (LIGO Academic Advisory Council) is responsible for the junior scientists within the LSC \cite{LAAC}. This includes supporting their education and recognition and generally their overall wellbeing within the collaboration, which implies a certain overlap with the work of the DEI committee.  
    \item Other possible points of contact are the LSC Ombudsperson
    (currently Lynn Cominsky --- \\ lynn.cominsky@ligo.org) who provides confidential, informal, independent, and neutral dispute resolution advisory services for all members of the LSC, and the LVC Allies who offer safe and confidential advice for all LVC members.
    \item At all LSC-Virgo-KAGRA collaboration meetings, DEI sessions and/or DEI-related talks and activities are offered.
\end{itemize}
The \href{https://dcc.ligo.org/LIGO-P1400033}{LSC Beginner’s Guide} helps new LSC members to get familiar with the LSC and provides up-to-date information on committees, wiki pages and websites and the responsible contacts \cite{LSC-beginners-guide}.

\subsubsection*{DEI in LISA}
The LISA consortium DEI committee leads DEI efforts for LISA. Specifically, this committee contributes to maintaining Consortium documents relevant to DEI, consults on Consortium leadership and committee appointments, and collaborates with DEI professionals and DEI groups within peer organizations such as the \href{https://astromdn.github.io/}{Multimessenger Diversity Network} (MDN) \cite{MDN}. Moreover, they organize DEI activities for Consortium meetings. Their policy documents can be found here: \href{https://wiki-lisa.in2p3.fr/pmwiki/uploads/Main/DEI_Charter.pdf}{DEI Charter} and \href{https://wiki-lisa.in2p3.fr/pmwiki/uploads/Main/DEI_Membership-Policy-Procedures.pdf}{DEI membership policy document} \cite{LISA-charter,LISA-policy}.

\printbibliography
\end{refsection}

\newpage
\begin{refsection}
\setcounter{figure}{0}
\setcounter{footnote}{0} 
\section{Well-being and Mental Health}
\begin{center}
    \textbf{Session chairs \& writers:}\\
    Arianna Renzini\footnote{LIGO  Laboratory, and Cahill astronomy department,  California  Institute  of  Technology,  Pasadena,  CA  91125,  USA}\\
    David Weir\footnote{ Department of Physics and Helsinki Institute of
  Physics,  P.O. Box 64, 00014 University of Helsinki, Finland}
\end{center}

\begin{center}
    \textbf{Invited speakers:}\\
    Juliana Perez\footnote{California  Institute  of  Technology,  Pasadena,  CA  91125,  USA}\\
\end{center}

\subsection*{Introduction}
Occupational therapist Juliana Perez gave a well-being and mental
health talk, focusing on problematic aspects of academic life. The
speaker provided us with detailed scientific descriptions of the
causes and symptoms of stress, how to distinguish good stress from bad
stress, and how to get out of unhealthy mindsets and into
healthy work habits. This was followed by an anonymous discussion/Q\&A
session via \href{https://flinga.fi/}{Flinga}, where
participants were able to post questions anonymously, vote for
questions they resonated with, and ultimately also intervene to
provide their view on the topics at hand. Overall, what emerged is
that there is a prevalent inability to cope with the societal expectations
for us as individuals, given the structure and mechanisms
of success in academia. There is also wide-spread impostor syndrome in
our community, which leads us to compare each other, and reflect upon
ourselves unfavourably.

\subsection*{Mental Health and Work-Life Balance}

In this section, we will explore the topics raised in Juliana Perez's
talk. A central component of any person's life is {\it stress}, and
their approach to handling stress sets the foundation of their
work-life balance and, ultimately, their mental health. We develop
this theme here, focusing in particular on the academic perspective,
and proposing useful methods for reducing the negative effects of
stress, and promoting a healthy relationship with one's work.

\subsubsection*{Stress: definition and symptoms}

\begin{figure}%[b]
    \centering
    \includegraphics[width=0.75\textwidth]{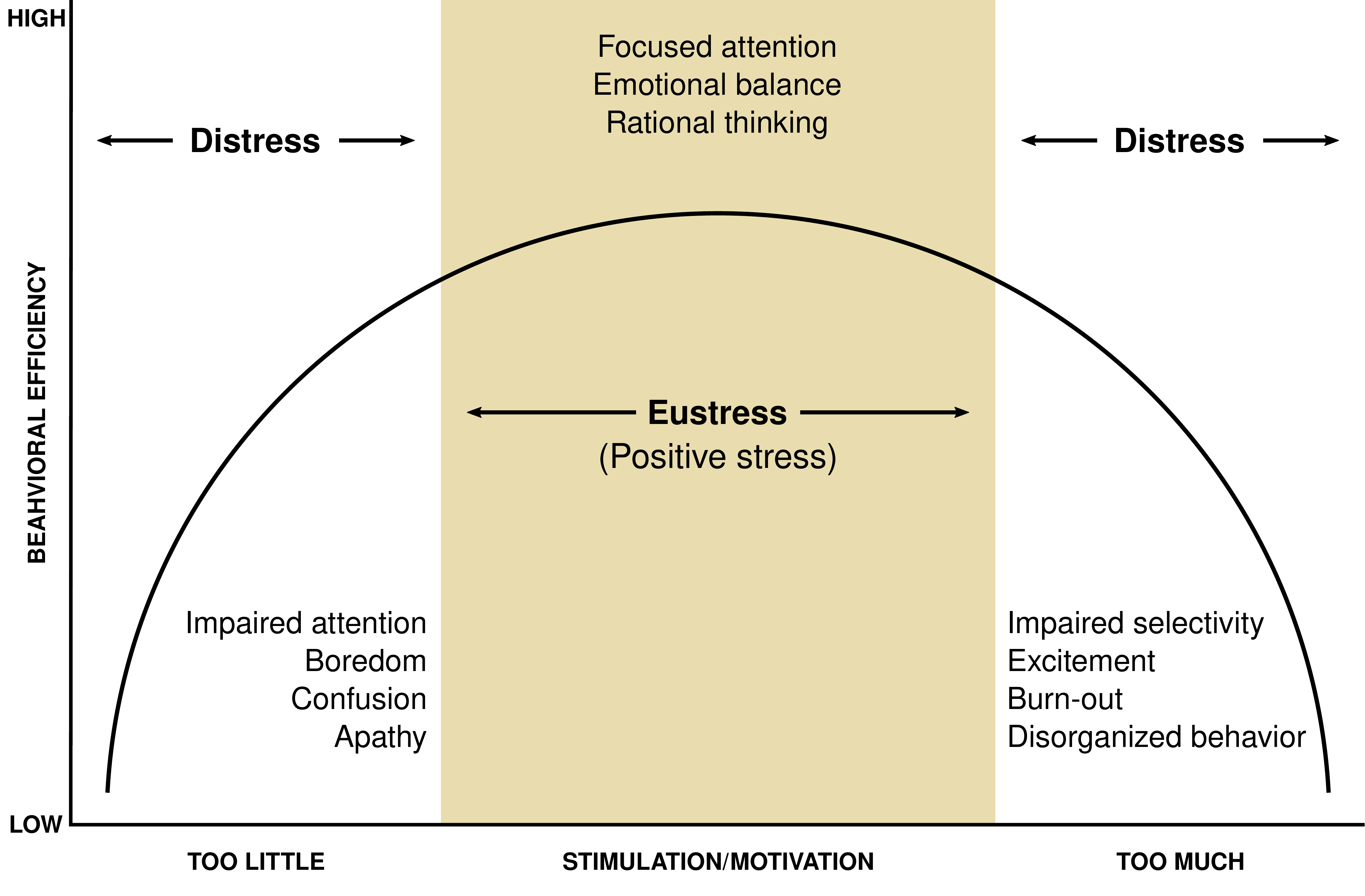}
    \caption{The Yerkes-Dodson Law of the stress spectrum~\cite{YerkesDodson1908}.}
    \label{fig:stress}
\end{figure}

To start, we categorize stress under two broad labels: {\it eustress} and {\it distress}.  Eustress promotes activity and productivity, and is often labelled ``good'' stress. Distress, on the other hand, impairs our productivity and is thus considered ``bad'' stress. Distress typically occurs when one is either over- or understimulated. This is formalised in the Yerkes-Dodson Law, see Figure~\ref{fig:stress}. Note that {\it everyone} experiences stress, yet as this representation shows, not all stress has a negative impact on one's mental health. To some degree, we require stress to function, and to be productive; the key to coping with stress lies in finding one's own balance on the scale in Figure~\ref{fig:stress}.

Stress can be caused by either {\it environmental stimuli},
perceived through sight, touch, sound, taste, smell, and
movement; or
{\it internal stimuli},
such as chronic or acute stress symptoms. When stress is
triggered in our body, the fight or flight response is activated,
interfering with
unnecessary systems such as digestion, sleep, the endocrine and
reproductive system, and executive functioning. Hence, one cannot be
under constant stress and be healthy, as one's body can not operate
properly in this state. Stress must then be relieved appropriately and
periodically, in order to ensure the body is under stress
only for short periods of time.

\paragraph{Stress symptoms.} Stress causes a wide range of symptoms. These can be defined as {\it emotional}, when someone who is stressed experiences fear, anger, nervousness, irritability, sadness, loneliness, hopelessness, moodiness, worry; {\it cognitive}, when someone who is stressed experiences repetitive thoughts, confusion, indecisiveness, lack of concentration, loss of confidence, poor memory; {\it physical}, when someone who is stressed feels sick, experience rapid breathing (hyperventilating), stomach aches, quickening heartbeat; they may also experience sweating, shaking, dizziness, muscle tension, headaches, fatigue, dry mouth, tightness in the chest; finally, these may be {\it behavioural}, when someone who is stressed experiences lack of sleep (insomnia), change in appetite, and may develop nervous habits, increased caffeine, alcohol, tobacco or/and drug consumption, and may start avoiding social and/or confrontational situations. 

\paragraph{Executive functioning.} Our work as physicists is
  mentally taxing, and it is crucial to explore how stress affects our
  ability to succeed in our work. The set of mental skills that enables us to work and be productive are referred to as {\it executive functions}. These include the following:
\begin{itemize}
    \item Response inhibition
    \item Working memory
    \item Emotional control
    \item Sustained attention
    \item Task initiation
    \item Planning and prioritizing
    \item Organisation
    \item Time management
    \item Goal-directed persistence
    \item Flexibility/ability to shift focus
    \item Metacognition, i.e. reflecting on one's own thinking.
\end{itemize}
Executive functioning ---
making use of our executive functions --- has a determining role in stress management and, ultimately, our ability
to perform research and contribute to the scientific community.

\subsubsection*{Managing stress}

Having recognised the possible symptoms of stress, and our set of
executive functions which can aid in dealing with stress, we can
tackle the topic of stress management.  It is possible to manage
distress and promote eustress. This may be done by identifying and
accommodating an adequate level of stimuli. This may be achieved by
engaging in either alerting or calming activities, as required.

General alerting activities include: light touch; chewing gum or
eating crunchy or spicy food; looking at reflective surfaces, textured
patterns and bright lights; listening to loud or fast music;
experiencing significant changes of temperature (cold or hot);
smelling particular scents such as cinnamon or peppermint; vestibular
inputs such as spinning, dancing, or rocking from side to side; and
proprioceptive inputs, meaning activities that pull or push against
joints such as running, climbing, and yoga.

General calming activities include: application of deep pressure, such
as massages, hugs, or a weighted blanket; slowly sipping a beverage or
sucking candy; keeping lights dimmed; listening to slow music;
experiencing warm temperatures; slow breathing activities which
encourage blowing, like blowing bubbles; smelling particular scents
such as lavender; slowly rocking back and forward; and reducing
clutter and distractions.

It is also possible to deal with stress by
addressing {\it maladaptive strategies}, i.e. identifying negative responses to
stress and trying to reduce them. This can be achieved also through the help of a specialist. Maladaptive strategies include:
\begin{itemize}
    \item Neglecting self self-care and hygiene.
    \item  Poor sleep hygiene –- insomnia, too much or too little sleep.
    \item  Poor eating habits –- not eating, not eating regularly, overeating.
    \item  Poor work-life balance –- overworking and not making time for participation in self self-care activities such as hobbies, exercise, visiting friends/family, getting outside.
    \item  Risky behaviors –- Substance Use Disorder, gambling, risky sexual behavior.
\end{itemize}
One can also find {\it adaptive strategies} to manage stress and promote wellbeing; these include
\begin{itemize}
    \item Creating a healthy “flow state” or being ``in the zone''.
    \item Regular exercise.
    \item Time management strategies, such as
      Pomodoro Technique, Covey’s Time Management Grid and using a planner regularly.
    \item Taking time for self-care and proper organization.
    \item Engaging in relaxing activities such as mindful meditation.
    \item Sleep hygiene -- winding down before bed and reducing distractions well before sleeping.
    \item Identifying peer support groups and professional support on and off campus.
    \item Creating study groups.
    \item Contribute to an
      open and safe environment.
\end{itemize}
Again, in some cases, one may attempt adaptive strategies successfully on one's own, however in other cases one should seek the help of a specialist.

As mentioned above, there are
  several {\it time-management} strategies that one may employ to be effective and manage time-related stress, such as deadline anxiety. One approach is called the {\it Eisenhower matrix}, which involves identifying one's important and unimportant activities, and categorizing them under one out of four labels: do it -- schedule it -- delegate it -- delete it. This allows one to consciously make decisions about what activities to engage in and when, helping the prioritization process. 
Another time-management strategy is the so-called {\it Pomodoro
  technique}, which involves identifying a single task to be done and
setting a timer to structure one's time to complete it
efficiently. For example, one might set a 25-minute timer to work on
the task with no distractions, take a five-minute break when the timer
rings, where one does not look at or think about the task, and repeat
until the goal is reached.
The recommendation is to repeat the 25-minute work/5-minute break framework up to four times, and then take a longer break.

\subsubsection*{Stages of professional development} 

\begin{figure}
    \centering
    \includegraphics[width = 0.75\textwidth]{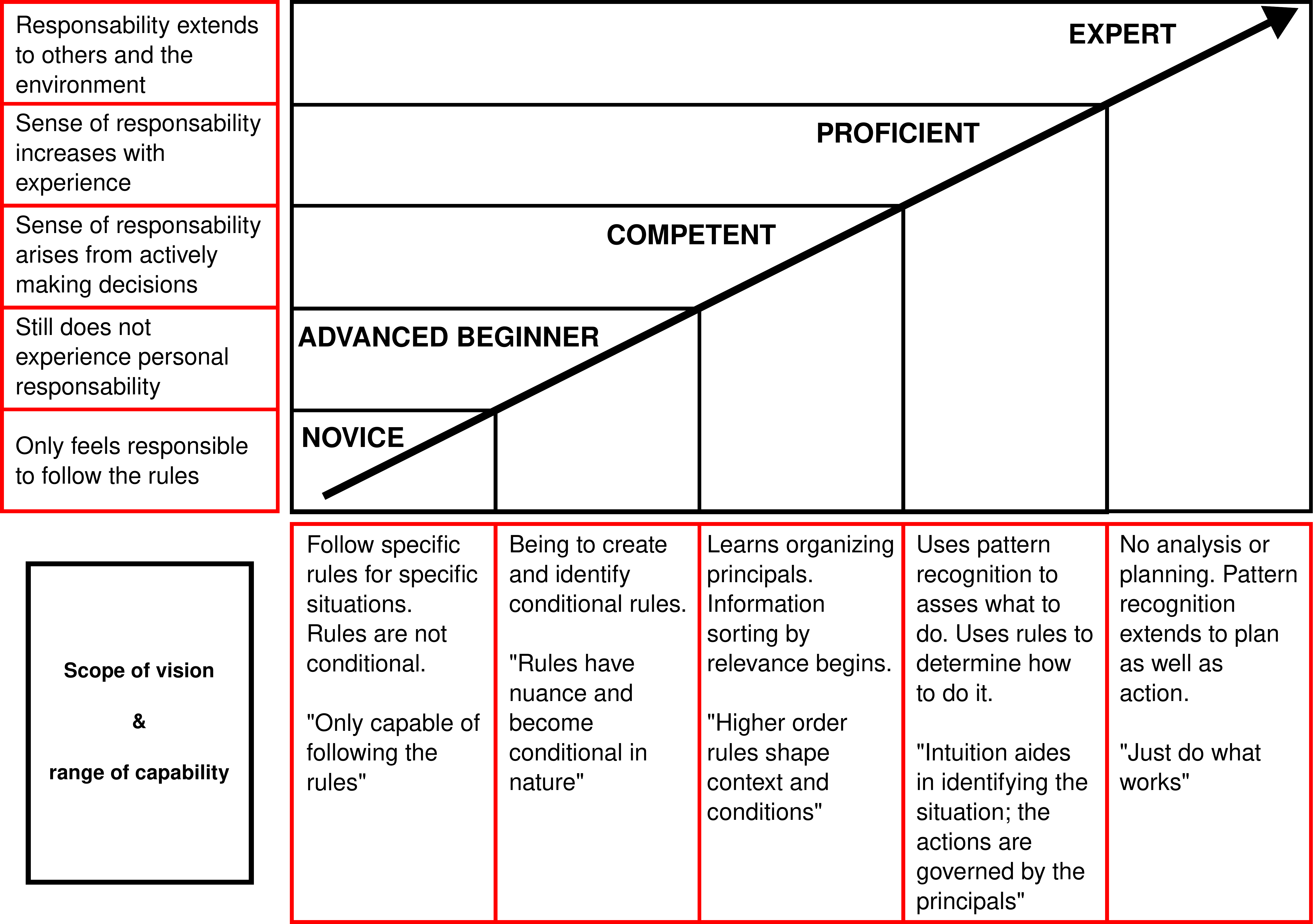}
    \caption{The Dreyfus model of skill acquisition~\cite{dreyfus1980}.}
    \label{fig:Dreyfus}
\end{figure}

Professional development and skill acquisition come in stages, hence
expecting to suddenly be ``the expert'' may contribute to stress,
perfectionism, and impostor syndrome. Perfectionism can hamper
success, as one sets exceedingly high standards and is
hypercritical of oneself, which ultimately leads to low self-esteem
and fear of being wrong and fear of evaluation. These
  feelings are particularly common among groups underrepresented in
  our field -- as well as those who do not subscribe to the false but
  widespread perception that it is necessary to be hypercompetitive to
  succeed in a science career. A useful reference for the stages of
professional development is the {\it Dreyfus model}, depicted in
Figure~\ref{fig:Dreyfus}.

In particular when comparing oneself with others, perfectionism may
lead to impostor syndrome, where one believes to be unworthy of the
position or role one is tasked with. 
Impostor syndrome is widely present in academia and disproportionately
affects high achieving people. Typical symptoms of impostor syndrome
are doubting one’s own abilities, feeling like a fraud, difficulty
accepting praise and feeling accomplished, and questioning whether
accolades and awards are deserved. A first step to overcoming impostor
syndrome is certainly recognising it, but beyond that it
may be useful to seek the help of a professional.

\subsection*{Discussion}
During and after Juliana's talk, we gathered questions and comments through a number of feedback channels. These included the GW\AE~CS Slack channel, the chat alongside the Teams video meeting itself, and an anonymous 'Flinga' board. Flinga allowed people to anonymously post questions or comments, as well as voting on other people's posts. We strongly recommend providing an anonymous means of contributing to sensitive discussions like this, as it removes the stigma and fear of speaking out about such topics. Meanwhile, the voting feature means topics which strike a chord with the audience can be discussed first. While there are very real, serious issues around wellbeing and mental health that only affect a small subset of our community, in the interests of limited time we could not cover all the topics raised through our feedback channels. The discussion here should, however, hopefully reflect all the points discussed.

On the other hand, there will be important challenges and issues which did not arise during our session; the discussion here is not intended to be exhaustive.
Regarding mental health, it was hard to avoid the reality of living in a pandemic and the detrimental effect it has had on our wellbeing. There have been more specific knock-on effects on mental health as a result of changed working practices. Most prominent of these is so-called Zoom fatigue, caused by spending a lot of time in online meetings. Some participants recommended turning off `self view' in video calls, so one cannot see one's own appearance or body language while talking - this can make the meeting feel more natural.

As much of our work involves programming and related tasks, there was a lively discussion of tools and platforms for pair programming - in other words, two (or more) people collaborating on the same code in real time. This can recreate more of a feeling of the shared experience of problem-solving than can sharing a screen in Zoom or Slack. Relatively easy-to-use options include Teletype, a plugin for the Atom editor; Amazon Web Services' Cloud9, and Codenvy.

Another pandemic-related issue was the feeling of powerlessness caused by being separated from family, friends and loved ones due to travel restrictions. Not only is the physical separation and inability to travel itself difficult, but as things return to normal in one country, one's peers may not appreciate the situation elsewhere - which could still be much worse, or deteriorating. It is therefore important to check in and empathise with our colleagues who may be lonely or worried about people living elsewhere.

For some, returning to the office as the pandemic eases may itself be a stressor. For those to whom the office is not a safe space, perhaps due to harassment or a toxic workplace culture, there is a need to disengage from the office to protect mental health.

A recurring theme was the need to make it easier to seek help when we are feeling down or stressed, particularly from professionals. Many people expressed a wish to destigmatise obtaining help or support, or finding ways to intervene to help others who appear burnt out or stressed.

Some of that stress and burnout likely comes from the toxic culture of overwork that often feels pervasive in our field. It is important to remember that there is no need to work long hours, or weekends, in order to have a successful scientific career. There is plenty of evidence to show that productivity does not increase significantly after working about 40 hours a week, and so working long hours is neither good for our science, nor for us. However, for many these long hours are perceived as necessary in order to meet frequent, tight deadlines. Others felt that they could not `keep up' with their peers without working long hours - a topic we will return to shortly.

The social pressure to work long hours is also related to several other aspects of work-related peer pressure or mismatched expectations around work-life balance. A particular challenge is finding a way to start or support a family while one or both parents are moving around from postdoc to postdoc. This is commonly known as the 'two-body problem', and may be worsened by pressure from family and friends maintaining expectations of settling down in one place soon after leaving higher education.
Even those who do not have children suffer from the constant moving around; it is hard to develop and maintain close friendships when moving every couple of years. Added to the challenge of making friends in a different culture and foreign language, it can be an alienating, isolating experience to be a postdoctoral researcher.

While many of us may wish to stay in academia, not everyone will be able to find a permanent job that both fits their personal constraints and scientific interests. Some participants in our event expressed concern about when they should start `worrying' about finding a plausible alternative to an academic career.

A final point of discussion around wellbeing was over the working environment. Many offices in comparable private sector jobs feature high quality, modern, ergonomic working areas, with well-designed spaces for formal and informal interactions. Meanwhile, a lot of academic offices are poorly laid out and not conducive to the teamwork and collaboration that characterises our field today.

There are other unhealthy working practices that come from sharing a space with others, besides the pressure to work long hours mentioned earlier. Some people reported comparing themselves, and their progress, unhelpfully with others around them; or that they lacked confidence in talking to their peers. There was a lack of relatable peer `role models' demonstrating best practices; most role models and mentors are at later career stages, so it is hard to infer from them what to do as a graduate student or postdoc.

The culture of overwork and peer pressure came up repeatedly in our session, with some people speaking of a `bragging culture' about working on weekends. Part of this may come from the lack of clear expectations from students or postdocs in our field; we are not issued with a list of tasks and goals, leading to a feeling that there is always much, much more to be done. Not talking about that intangible mountain of tasks inevitably leads to stress, which could be reduced by better target setting, and coordination with supervisors and collaborators.

More generally, there was a feeling that power structures, hierarchy and the feeling of distant, out-of-touch supervisors made it hard to negotiate what was needed and how to achieve it - or to seek help and advice both about work and about avoiding some of the negative consequences discussed here. A broader debate about how to flatten the power hierarchy in academia would help to alleviate many of the issues discussed here.

\printbibliography
\end{refsection}

\newpage
\begin{refsection}
\setcounter{figure}{0}
\setcounter{footnote}{0} 
\section{Outreach and Funding Opportunities}
\begin{center}
    \textbf{Session chairs \& writers:}\\
    Elisa Maggio\footnote{Dipartimento di Fisica, ``Sapienza'' Universit\`a di Roma \& Sezione INFN Roma1, Piazzale
    Aldo Moro 5, 00185, Roma, Italy}\textsuperscript{,}\footnote{Max-Planck-Institut f\"ur Gravitationsphysik, Albert-Einstein-Institut, 
    Am M\"uhlenberg 1, 14476 Potsdam-Golm, Germany}\\
    Angelo Ricciardone\footnote{Dipartimento di Fisica e Astronomia ``G. Galilei",
    Universit\`a degli Studi di Padova, via Marzolo 8, I-35131 Padova, Italy \& INFN, Sezione di Padova,
    via Marzolo 8, I-35131 Padova, Italy}
\end{center}

\begin{center}
    \textbf{Invited speakers:}\\
    Davide Gerosa\footnote{Dipartimento di Fisica ``G. Occhialini'', Universit\`a degli Studi di Milano-Bicocca, Piazza della Scienza 3, 20126 Milano, Italy \& INFN, Sezione di Milano-Bicocca, Piazza della Scienza 3, 20126 Milano, Italy}\\
    Martin Hendry\footnote{SUPA, School of Physics and Astronomy, University of Glasgow, Glasgow G12 8QQ, United Kingdom.}
\end{center}

The aim of this session of the workshop was to have a presentation about outreach activity for gravitational waves (GWs) and one about early career scientists (ECS) funding opportunities and how to get them. The session was chaired by Elisa Maggio and Angelo Ricciardone. The first talk was given by Martin Hendry, who is a Professor at the University of Glasgow, chair of the LSC Communications and Education Division and co-chair of the Advocacy and Outreach Group of LISA. The second talk was given by Davide Gerosa, who was a Lecturer at the University of Birmingham, now Associate Professor in Milano Bicocca and a recent ERC Starting Grant awardee.

\subsection*{Outreach for GWs}

During the first talk, Martin presented outreach activities as a tool to link the scientific research that scientists do to popular culture. Martin stressed how impressive the achievements in our field are, and how many times the science that we do sounds like science fiction. 

Martin claims that one of the reasons for doing outreach is the energy that we capture from tapping into the excitement of the public. It represents the collaborative effort of a lot of people. 
Outreach has an impact also on research funding, since most funded schemes look for researchers to demonstrate how their research reaches public audiences and has an impact on everyday life. Among the reasons for doing outreach Martin mentions: 1) to have fun, 2) to generate impact, 3) to share research, 4) to develop skills, 5) to inspire learning, 6) to improve research, 7) to raise aspirations, 8) to build trust and inform policy, 9) for public accountability, 10) to learn from others, and last but not least to improve relationships with communities.

\begin{figure}[t]
\centering
\includegraphics[scale=0.25]{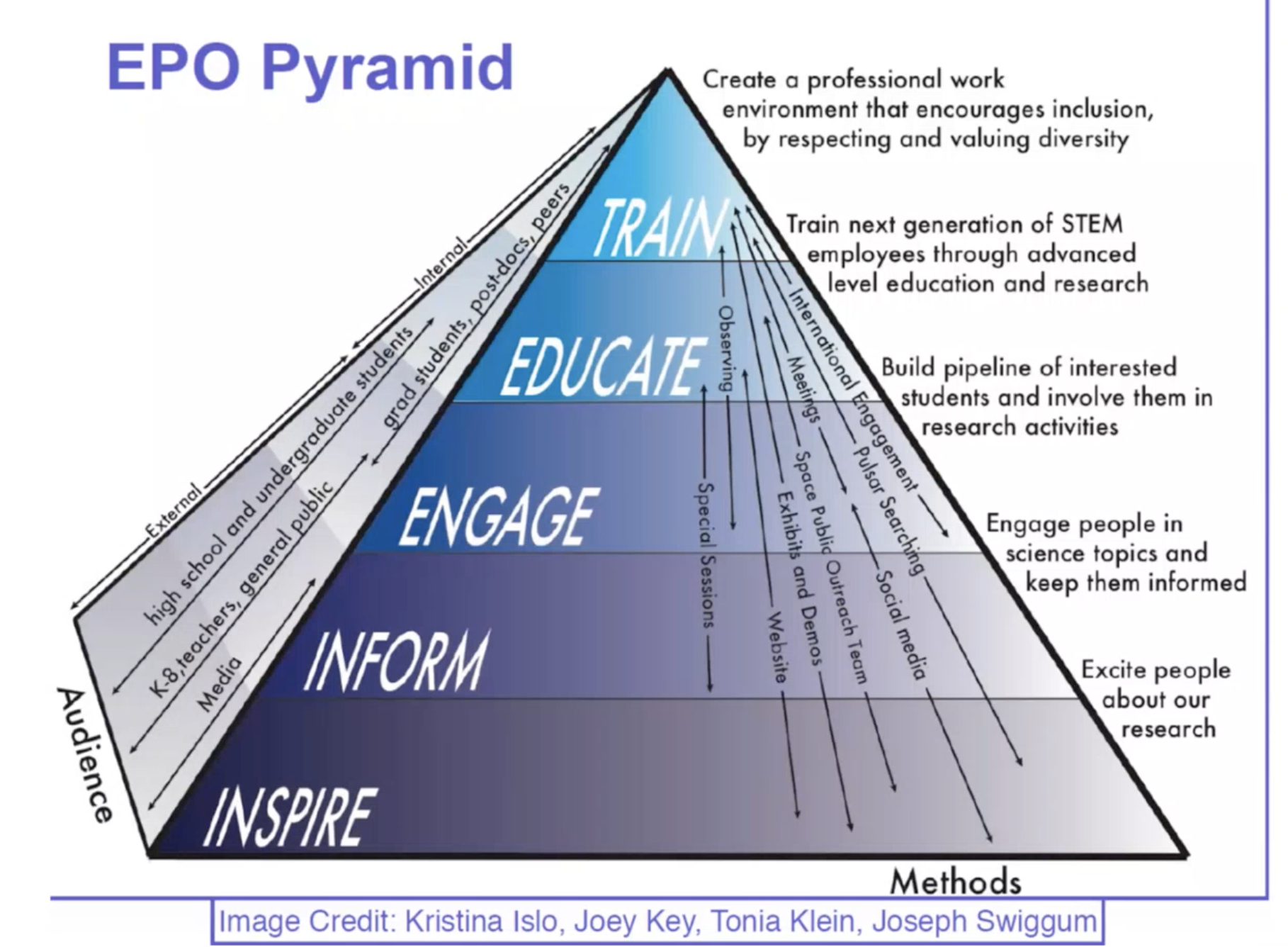}
\caption{\it Educational and Public Outreach (EPO) pyramid to build a well-trained scientific community through outreach. Credit: K. Islo, J. Key, T. Klein, J. Swiggum (NANOGrav).}
\label{fig:EPOpyramid}
\end{figure}

There are different levels in which outreach activities can operate: as shown in Fig.~\ref{fig:EPOpyramid}, outreach can operate mainly through informing, engaging and educating. It is important to consider that the audience can become the next generation of scientists. Also for non-scientists outreach can be a way to have a better appreciation of why research is important.
The Educational and Public Outreach (EPO) pyramid in Fig.~\ref{fig:EPOpyramid} shows the important and necessary steps to build a well-trained scientific community through outreach efforts. The left side of the pyramid shows the different audiences, which range from the general public to students at different levels, postdocs and peers. The pyramid also shows the common methods that can be used in outreach: from social media to public exhibits, Space Public Outreach Team (SPOT) presentations, and special sessions during scientific meetings.
The vertical dimension describes the goals of the outreach activity: from inspiring a broad range of people to gradually focusing the effort to training students and junior researchers.

Martin mentioned several activities that have been organized to involve people. These include:  developing materials that can be easily adapted to different audiences; virtual exhibitions (even if they are not as engaging as face-to-face sessions); science summaries for the general public, i.e., journalists, teachers, high school students, and their translation into as many languages as possible; activities on social media, e.g., about GW alerts. The pandemic situation has had a major impact on most of these outreach activities, forcing almost all of them to move online. This has made it important to use reliable and easy to use online platforms to better attract (and retain) audiences. While some materials have been created by professional artists and designers, most of the resources developed by the GW community -- e.g. to accompany public announcements of LIGO and Virgo detections -- have been created by the scientists themselves.

\subsubsection*{Suggestions for ECSs interested in outreach}

Martin highlighted some important strategies for delivering effective outreach. These include using lots of analogies and pop culture references and focussing not just on the science results but on the scientific process that produces them. Sometimes the science is not described perfectly in order to keep it simple for a general audience, but it is much more important to keep people engaged.

About learning how to do outreach, it is important to have mentors who are very supportive so that ECSs can learn by example. Nowadays  there is also some formal training available (e.g., the recently-formed IGrav~\cite{IGRAV} is planning to organize communication trainings open to the entire GW community). A good practical training for grad students could be to contribute to the website “Astrobites”~\cite{astrobites} that summarizes papers in the arXiv~\cite{arXiv} and always looks for new writers.

Good practice is to try to link in with existing outreach activities, to avoid re-inventing the wheel and improve efficiency. It is also important that the collaborations in which ECS work recognize the value of outreach as a core activity and not as something that is done in free time. Outreach may also have a crucial role when making fellowship applications, in fact many applications ask for a statement on outreach and impact plans. Experience in outreach can reinforce and amplify the value of the research activity of ECSs.

On the importance of social media for outreach, Martin recognizes that it is a very effective way to reach a large and diverse audience. For example public alerts of GWs were widely shared on social media, increasing the audience well beyond other astronomers -- even though this was not essential for follow-up of GW alerts. However, it is also important to have in mind  other communication channels because not everyone is active on social media.
Many people promote new papers on Twitter and there is an open discussion about the impact of this in the scientific community~\cite{10.1371/journal.pone.0229446,LUC2021296}. 

A final comment is on the possibility to take on outreach as a job: there are many science centers that have such positions, and will be looking for well-trained and experienced researchers. Many universities also offer training programs on communication for ECSs. In the United Kingdom, for example, the Institute of Physics organizes a high-profile competition rewarding excellence in outreach and science communication \cite{3mw}.

\subsubsection*{Poll about outreach}

\begin{figure}[t]
\centering
\includegraphics[scale=0.3]{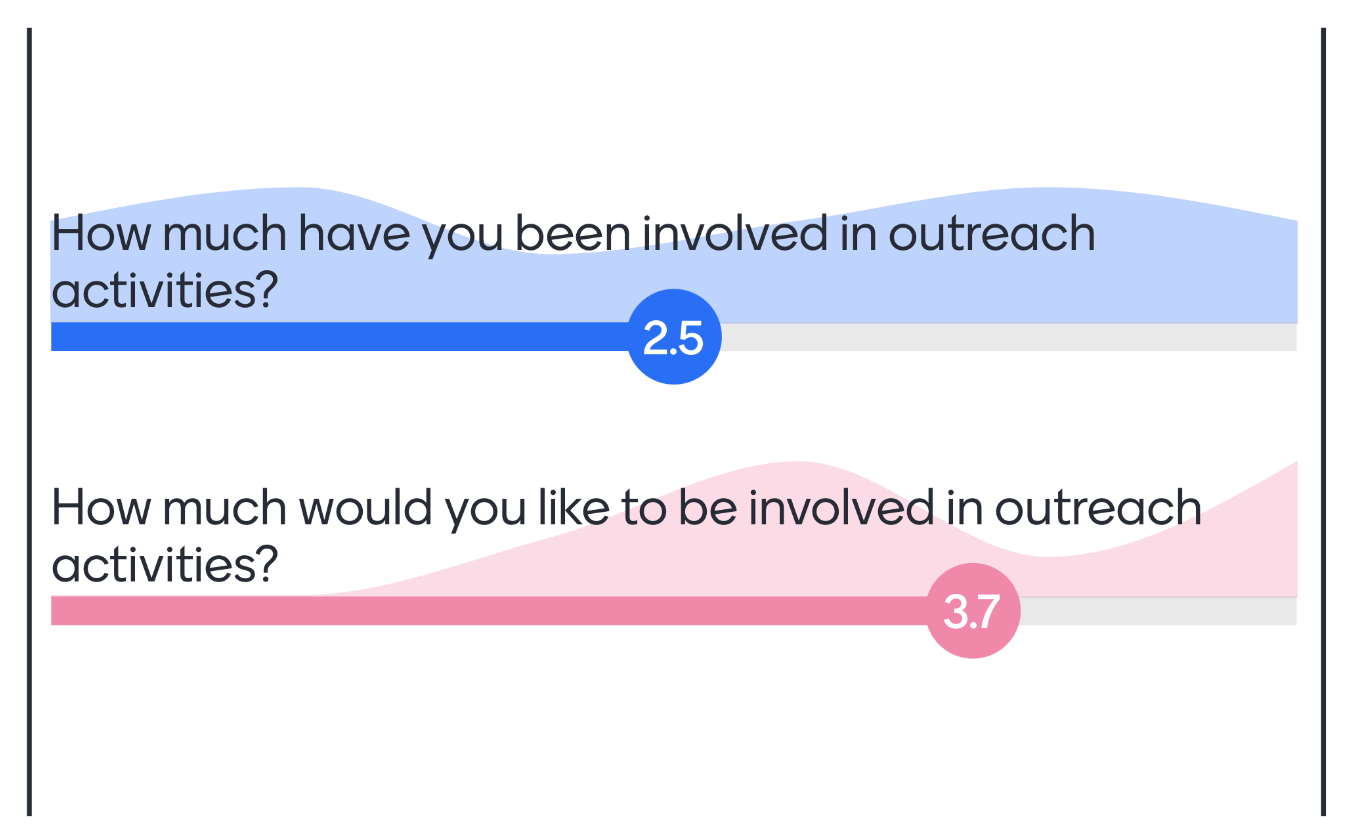}
\caption{\it Results of a poll about outreach completed by twenty participants of the GWÆCS workshop.}
\label{fig:outreachpoll}
\end{figure}
Twenty participants of the workshop participated in a poll about outreach. The audience was asked how much they have been involved in outreach activities and how much they would like to be involved in outreach activies in a range from one to five. 
As shown in Fig.~\ref{fig:outreachpoll}, the participation of the audience in outreach activities is diverse and the distribution is almost flat with an average of $2.5/5$. Overall ECSs would like to be involved more in outreach activities with an average of $3.7/5$.

\subsection*{ECSs funding opportunities}

The second talk of the session was presented by  Davide and was titled ``Transitioning to research independence in gravitational-wave astronomy (my own view on postdoc and grant applications).'' After having obtained a PhD in Physics or in Astrophysics, ECSs should ask themselves whether they would like to apply for a postdoctoral position or not. It is a choice and not a one-way street. The same argument applies after having completed one or two postdocs. It is a choice of the ECS whether they would like to become a professor. Because of the personal experience of the speaker (Davide) as well as the nature of the GW{\AE}CS  workshop, the discussion was restricted to academic jobs. It must be stressed that there are many, intellectually stimulating career paths, and academia is only one of them. 

\subsubsection*{Some numbers}

Concerning the number of applicants in the astronomy job market, let us analyze data collected in the United Kingdom as a test case. Extensive survey from the Royal Astronomical Society \cite{RAScareer,10.1093/astrogeo/atx211} found that about a third to half of the finishing astronomy PhD students in the country find a post-doctoral appointment in the field. But only a fifth of those that graduate will achieve a stable academic career. Considering that about $\sim 50\%$ of the graduating PhD students choose to leave academia, is not unreasonable to say that about a third of the graduates that want a faculty job get one. 

It is important to stress that this estimate is only indicative and changes depending on country,  institution, social background, gender, etc., which are certainly contributing factors. 

Outside academia there is a wide range of views on the hiring of applicants. Some employers prefer young people, whereas other ones evaluate the range of skills of postdocs. Usually postdocs have strong curriculum vitae (CVs) because they are trained on skills that can be applied to a wide range of activities.

\subsubsection*{Postdoc vs fellowship}

Post-PhD academic jobs can be divided into two broad categories.

\begin{enumerate}

\item A \emph{postdoc} is tied to the grant of a Principal Investigator (PI). The postdoc joins the team lead by the PI to work on a specific project. Typically, these projects are of extremely high profile (grants are hard to get!). An advantage of this kind of positions is that the ECS works in a well developed group with a strong, key idea. This is the perfect opportunity to publish many papers in a short period of time.

\item A \emph{fellowship} is instead a grant awarded to the ECS to work in a host institution.  Fellowship appointments are typically considered more prestigious than a standard postdoc because they demonstrate independence, which is a key prerequisite for a faculty job. There are drawbacks. The ECS is essentially on their own and thus needs to know what to do in research and lead their own investigations (in terms of priorities, collaborations, etc.).

\end{enumerate}

The two processes for how the jobs are awarded are a bit different.
For postdoc applications, a PI looks for a specific profile. It is important for an applicant to find out who the PI is and write a proposal tailored to them because they are going to decide. The applicant does not need to convince the PI that their research field is worth pursuing because the project is already defined. The ``one proposal to rule them all'' strategy does not work because each  opening has a specific research project. 

For fellowship applications instead, it is important to show leadership potential. First, the applicant needs to find a host institution. This is usually not a major issue, as the vast majority of the institutions are eager to attract applicants, so a suggestion for the applicant is to not be shy in asking (there are some exceptions, as some schemes only allow universities to support a limited number of applicants; in those cases, internal selection processes are typically in place). A fellowship proposal is evaluated by a panel composed of both experts and non-experts in the research field of the applicant. 
In many cases, proposals are first sent to external reviewers and then lands on the desk of the panel members. The program officers designate a primary and one or more secondary reviewers for each proposal. The primary reviewer needs to read that specific proposal more carefully and kickstart the discussion with a presentation highlighting strong and weak points. The secondary reviewers can then bring forward additional points for consideration. This is followed by a panel discussion where a ranking is assigned. The crucial point is to convince the primary reviewer, which can be assumed to be reasonably close to the research field of the applicant, that one's proposal is worth funding. A negative report from the primary reviewer can be enough to kill a proposal. In the panel discussion, it is also important to convince the non-experts that the proposed research is useful for their science. In the best proposals are present specific activities that are well inserted into a broader horizon.

A suggestion for ECSs is to accept the participation in panels if they are invited. Although   sitting in a grant review panel is a lot of work (like 10 or 20 times more than refereeing a paper),  it is extremely useful to see how the application process works from the inside.

\subsubsection*{Faculty applications}

After some postdoc years, one may decide to apply for a faculty position.  ECSs are ofter invited to apply for faculty positions, but this is not a guarantee that they will be hired, or even that that they are prime candidates. Multiple people are invited, sometimes because the department needs to show  the University that they can attract several candidates. 
It is important for the applicant to know about the environment and the priorities of the institution they are applying for. Higher-education institutions can be research or teaching oriented, thus the applicant need to strengthen different sides of their portfolio according to what they want to hear.
Moreover, it is useful to know how broad the search is for that position (are they specifically looking for someone in GWs? Or is it a broader search in Physics or even all of the sciences?). With a little effort, it is possible to understand whether they have to convince a broader panel that GW physics is interesting, or conversely avoid wasting time in needless introductory material if the search is very specific.

Often the applicant needs to provide a teaching statement. A general suggestion is to make it personal in order to try to stand out from other statements. Some institutions also ask for a diversity statement. These are both difficult to write, as it is hard to be original. For research-intensive environments, they are not the most crucial part of an application.

When the applicant is invited for an interview, they should try to give a broad overview of the field/subfield and provide an argument why hiring them is important for the long-term goals of department (say on a scale of 20 years). A faculty job talk is not a normal talk where one presents the last paper. 

Some of the not-so-trivial questions from Davide's experience: 
\begin{itemize}
    \item We hire you and give you a PhD student starting immediately. What's their project going to be? [This is to check if you have scientific ideas for others, not just yourself].
    \item Who will you be trying to hire as postdocs (with names)? [This is to check if you know your community].
    \item What grants will you be applying for your first years? [This is to check you know the funding landscape and won't be wasting time looking around].
    \item What classes are you or are you not willing to teach? [Be honest (do not say you can do quantum field theory if you have never studied it yourself!) but at the same time show flexibility and willingness]. 
\end{itemize}

\subsubsection*{CVs in big collaborations}

Standing out in big collaborations is  challenging. The reason is that all the CVs of people who joined the collaboration at the same time will look very similar (same publications, more or less same working group positions, etc.). It can be very hard  for a reviewer to pinpoint the applicant's actual contribution and thus their leadership potential. Unless the employer is in the same collaboration, reference letters are often the only way to tell. A general suggestion for ECSs in collaborations is try to be self-critical, distinguish papers in which they contributed substantially and indicate them clearly in their publication list (and be ready to stand questions on those findings!). Another suggestion is to be specific about the collaboration activities explaining what leadership/responsibility the ECS had in a working group. It is important to make clear the contribution of the ECS and give a context to the meaning of a quoted leadership position. 

The ECS should also consider if they are willing to leave the collaboration to get the job they are applying for (if the employer is not in LISA or LIGO/Virgo, they might not to accept a situation where a postdoc cannot tell their supervisor what they are working on because of collaboration rules).

\subsubsection*{Various tips (not a must have list!)}

This is a list with various job-hunting tips, in no particular order.

\begin{itemize}
    \item \textbf{Read it quickly.} The applicant should consider that their application pack will be read in $5-10$ minutes. For this reason, it is important for the applicant to stress key points, e.g., in boldface or in boxed paragraphs. Ask a friend/supervisor in the field to read your application in 5 minutes and tell you what are the key points. Do they agree with what you want people to remember?
    \item \textbf{Do ECSs need papers?} Yes. However, there are not strict rankings or hard cuts (see e.g. Ref.~\cite{2019PASP..131a4501P} for an analysis of recent NASA Fellowship awards). An advice to show independence is to publish without one's PhD supervisor. The impact factor of the journal is typically not important for postdocs applications, whereas it could be more important for faculty applications because in some countries those are used in the department's evaluation. ``In preparation'' papers should not be included because they can be perceived as an attempt to inflate the list. Separate (or even avoid) proceedings, collaboration papers, etc., but include submitted/arXiv papers (in GWs we tend to post to the arXiv only pieces of work that are essentially complete).
    \item \textbf{Are reference letters important?} They are very important in all cases, but especially for candidates in big collaborations. A general suggestion is to ask reference letters to well known professors only if they know the ECS. A reference letter from a postdoc is ok but only one out of three. Employers are human and is thus likely that they will pay more attention if they know the letter writer and have a good opinion of them. Ask for reference letters well in advance (especially the first one to a new referee!), because less time will translate into a weaker letter.
    \item \textbf{Should graduate students stay for one more PhD year, if possible?}  It is a very personal decision. It is important to consider that the academic clock typically starts when ECSs graduate. Many fellowship and grant applications have as eligibility requirements 
    that ECSs are within a certain number of years after their PhD.
    \item \textbf{Should ECSs have a website?} Yes, especially as a new faculty to attract students and postdocs. About science promotion on social media, Davide has a very limited experience on it and declined to comment.
\end{itemize}

\subsubsection*{The job cycle}

Some useful websites where ECSs can look for jobs are:
\begin{itemize} 
\item hyperspace is specific to gravity/GWs~\cite{hyperspace};
\item the AAS job register is the key venue in Astrophysics~\cite{jobregister};
\item INSPIRE is the key venue in Theoretical Physics~\cite{inspirehep};
\item a list of research fellowships managed by the LISA Early Career Scientists group (LECS)~\cite{LECS}.
\end{itemize}

\noindent The typical timeline for the job cycle consists of: 
\begin{itemize}
\item new posts in August-October, 
\item deadlines in November-December, 
\item offers in January-February,
\item starting date in October.
\end{itemize}
This timeline is indicative and very country specific.

There is a resolution from the American Astronomical Society (AAS) asking to employers to not force ECSs to accept or decline a position before February $15\rm{th}$ each year \cite{AASres}. This is to protect candidates, so that they can have the full range of options in front of them before deciding. What happens is that on that day all the institutions know the answer from their first-choice candidates, second-choice candidates are contacted, etc. The same thing happens in the theoretical physics community, but on January $5\rm{th}$ \cite{tppost}. GW science is in between astronomy and theoretical physics and our deadlines often overlap. Moreover some wider fellowships, i.e., Marie Skłodowska-Curie Individual Fellowships, have different timelines.

\subsubsection*{Final remarks}

ECSs are encouraged to ask for help in job applications.
Typically other ECSs in the same field have already applied to the same fellowship a few years before. It is useful to send grant proposals to collaborators for comments and ask them to be critical (but always remember people are busy and you might not get feedback at all). When applicable, exploit the grant office of the host institution for all the budgetary and standard forms of the application.

In Davide's view, employers look for ideas more than abilities. If told what to do, many PhDs could write codes and papers. The key talent is to come up with \emph{the idea} for a strong paper.

Employers also look for versatility. Davide's suggestion is to explore a wide range of topics in GWs and get into new research strands without working on the same thing over and over. The strategy ``I will do what I am doing but better'' does not work in either postdoc applications or grant proposals. The ECS should have strong points and previous work that they would like to complete, but always with an element of exploration.

\subsubsection*{Poll about funding opportunities}

\begin{figure}[th]
\centering
\includegraphics[scale=0.3]{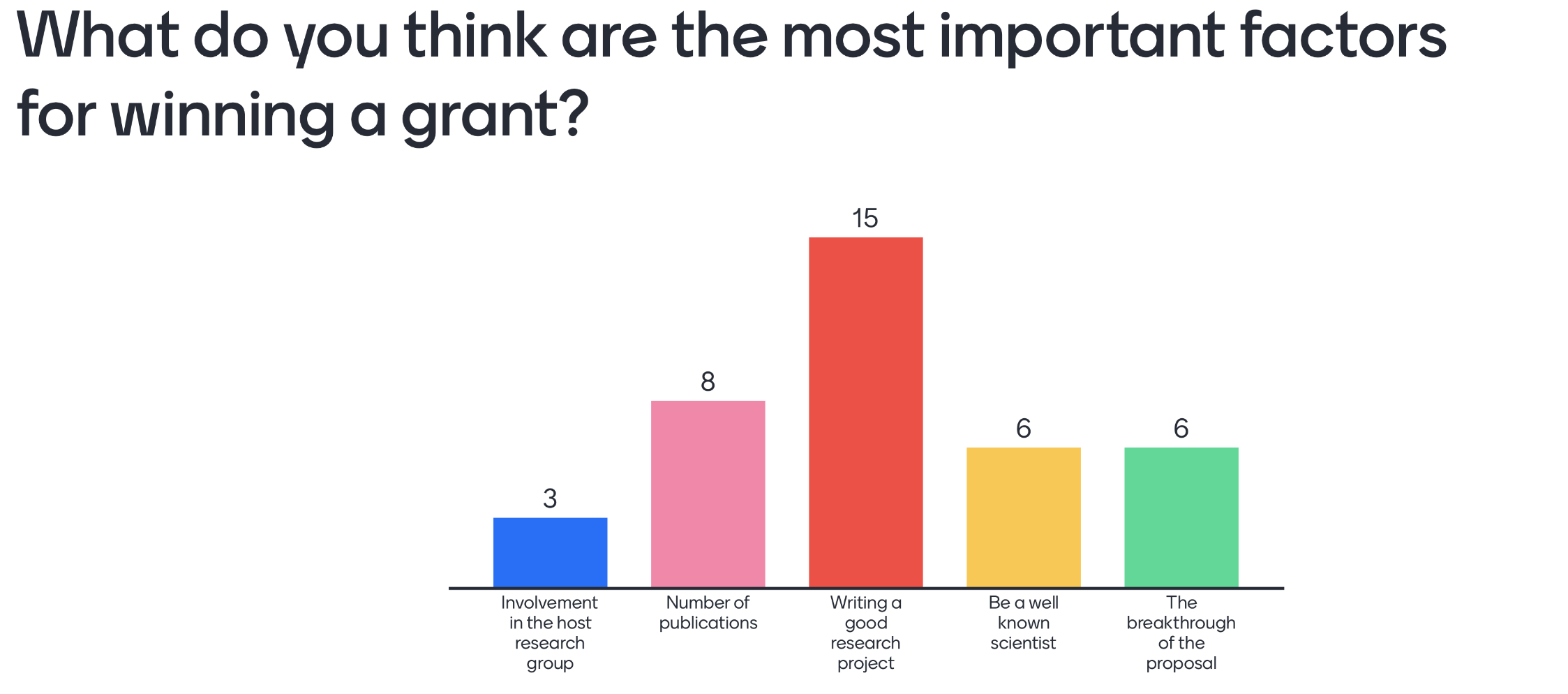}
\caption{\it Results of a poll about funding opportunities completed by twenty participants of the GWÆCS workshop.}
\label{fig:grantpoll}
\end{figure}
Twenty participants of the workshop participated in a poll about funding opportunities. The audience was asked what they believe are the most important factors for winning a grant among five options. 
As shown in Fig.~\ref{fig:grantpoll}, the vast majority of the audience thinks that writing a good research project is the most important factor, with $15/20$ votes. The second most voted option was the number of publications, with $8/20$ votes. Other options were “be a well known scientist” ($6/20$ votes), “the breakthrough of the proposal” ($6/20$ votes) and the “involvement in the host research group” ($3/20$ votes).

\subsection*{Future activities for ECSs}

The audience was asked what would they recommend as future activities for ECSs. In the following there is a list of the replies:
\begin{itemize}
    \item More events to support ECSs in job applications inside and outside academia.
    \item More events like the GWÆCS workshop.
    \item Round table discussion with senior scientists about how to write a research project
    \item Seminars with people that have been part of grant panels.
    \item Trainings for early career faculty about how to supervise students.
    \item Hands-on tutorials on specific science topics.
\end{itemize}

\begin{figure}[th]
\centering
\includegraphics[scale=0.3]{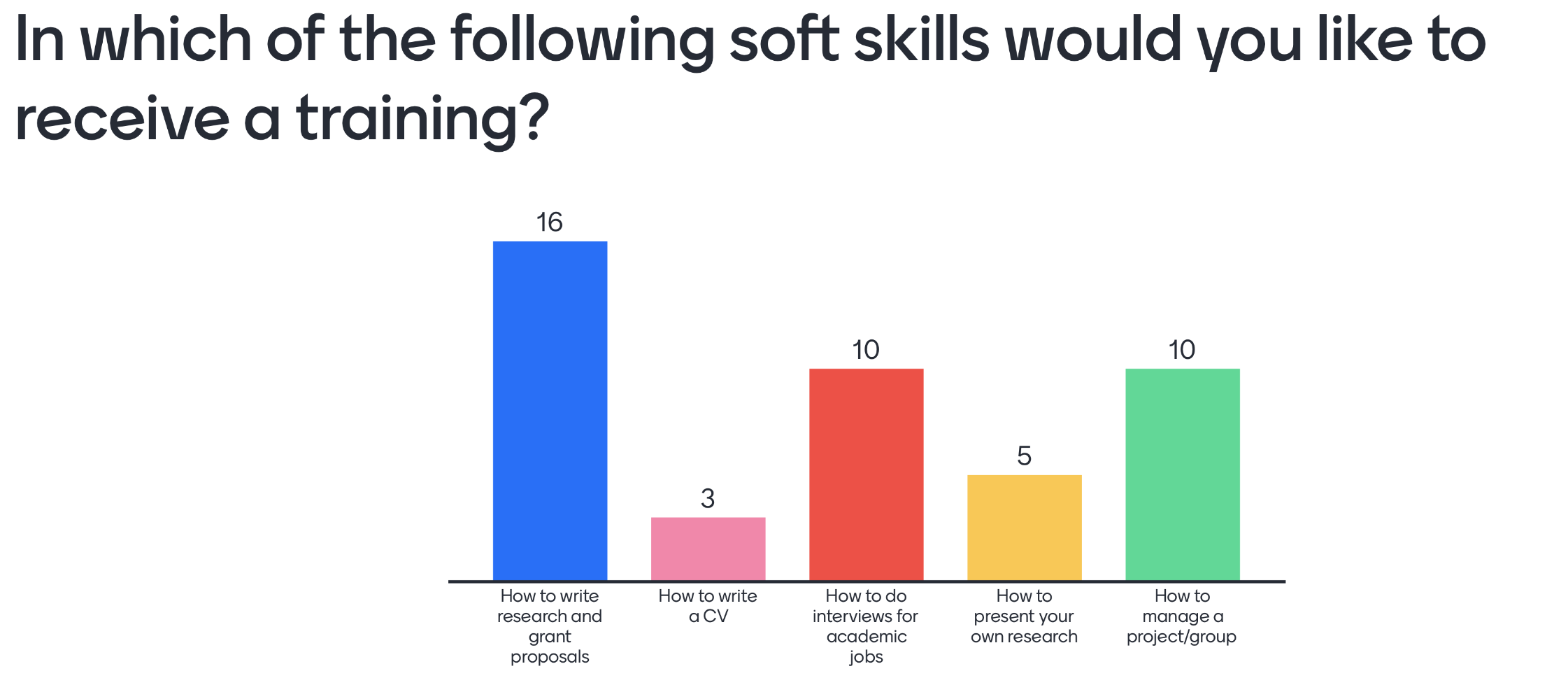}
\caption{\it Results of a poll about soft skills trainings completed by seventeen participants of the GWÆCS workshop.}
\label{fig:softskillspoll}
\end{figure}

Furthermore the audience was asked in which soft skills they would like to receive a training among five options. Seventeen people participated in the survey whose results are shown in Fig.~\ref{fig:softskillspoll}. Almost all the participants would like to receive a training on how to write research and grant proposals, with $16/17$ votes. A majority of participants is interested in trainings on how to do interviews for academic jobs ($10/17$ votes) and how to manage a project/group ($10/17$ votes). Other options were “how to present your own research” ($5/10$ votes) and “how to write a CV” ($3/10$ votes).

\printbibliography
\end{refsection}

\newpage
\part{Science Sessions}

\newpage
\begin{refsection}
\setcounter{figure}{0}
\setcounter{footnote}{0} 
\section{Gravitational Wave Scientist Toolbox}
\begin{center}
    \textbf{Session chairs \& writers:}\\
    Lorenzo Speri\footnote{Max-Planck-Institut f\"ur Gravitationsphysik, Albert-Einstein-Institut, 
Am M\"uhlenberg 1, 14476 Potsdam-Golm, Germany}\\
    Nikolaos Karnesis\footnote{Department of Physics, Aristotle University of Thessaloniki, Thessaloniki 54124, Greece}
\end{center}

\begin{center}
    \textbf{Invited speakers:}\\
    Jonathan Gair\footnote{Max-Planck-Institut f\"ur Gravitationsphysik, Albert-Einstein-Institut, 
Am M\"uhlenberg 1, 14476 Potsdam-Golm, Germany}\\
    Daniel Kennefick\footnote{Department of Physics, University of Arkansas, Fayetteville, Arkansas 72701, USA}
\end{center}

\newcommand{\ls}[1]{\textcolor{red}{Lorenzo: #1} }
\newcommand{\nk}[1]{\textcolor{blue}{Nikos: #1} }
\newcommand{\jg}[1]{\textcolor{green}{Jon: #1} }

%-------------------------------------------------------------------------------------------------

Gravitational wave observations offer unique opportunities to probe the fundamental properties of compact objects, to explore gravity in nonlinear, dynamical regimes, and to advance our understanding of the history of the Universe.
However, as 
in all measurements, the 
detector outputs contain not only the gravitational wave signal, but also an instrument noise component. Therefore, to fully exploit the potential of GW observations it is crucial to correctly understand 
the properties of the instrumental noise and disentangle it from the GW signal. Moreover, future GW detectors will be sensitive enough to measure stochastic signals from astrophysical, and possibly cosmological origin. These types of signals will add more complexity to the data analysis practices, because they will manifest in the detector bands effectively as another noise source.

There are typically four main stages in the analysis of gravitational wave signals: pre-processing, search, detection and parameter estimation. These different stages are not independent from each other and are influenced by 
the noise properties. The first step refers to the pre-processing of the data before being input to the statistical analysis pipelines. This includes the completion of all calibration and data quality checks. Assuming that we have an adequate understanding of the instrument noise, we begin the search phase. To search for a signal, it is necessary to assume (or adopt) a prior belief on the kind of signal present in the data stream. Depending on the GW detector design and capabilities, we first identify the types of sources that can %be 
actually be present in the given range of frequencies that the detector is sensitive to, and then search for all source types.

Once possible candidate signals are identified in the data, these are characterised through a more targeted parameter estimation analysis using accurate models; we infer the parameters of the system or of the process which is producing it. At the same time, it is necessary to understand how likely it is that noise reproduces such signal.

Once we have estimated the parameters we can finally use them for addressing scientific questions!

%%%%%%%%%%%%%%%%%%%%%%%%%%%%%%%%%%%%%%%%%%%%%%%%%%%%%%%%%%%%%%%%%%%%%%%%%
\subsection*{Likelihood, Posterior, Fisher Information Matrix, Matched Filtering}
%%%%%%%%%%%%%%%%%%%%%%%%%%%%%%%%%%%%%%%%%%%%%%%%%%%%%%%%%%%%%%%%%%%%%%%%%

Detection and parameter estimation of gravitational wave signals employs a wide range of techniques used to identify and characterize a signal $h(t)$ (hopefully) present in the output, $s(t)$, of a gravitational wave detector.
We review some basic concepts of gravitational wave data analysis that are widely used in the literature \cite{gairMakingSenseData}.
We assume that the output of a gravitational wave detector is composed of a signal $h(t; \vb*{\lambda} )$, dependent on the source parameters, $\vb*{\lambda}$, and instrumental noise $n(t)$:
\begin{equation}
    s(t) = h(t ;\vb*{\lambda} ) + n(t). \qquad \qquad 
\end{equation}
The probability $p(s|h)$ of measuring $s(t)$ given a particular signal $h(t; \vb*{\lambda})$ is the same as the probability $p(s' = s-h| 0)$ of measuring $s- h$ given that the signal $h$ is not present in $s'$, i.e. $p(s' = s-h| 0)$  is the probability that the noise realization would take the value $n = s-h$ \cite{finnDetectionMeasurementGravitational1992}.
Notice that this is possible because of our assumption that the output is a linear combination of signal plus noise.

\paragraph{ The likelihood:} describes the probability of observing a particular detector output, as a function of the parameters of the gravitational wave signal model. Writing this down relies on a model of how the data is related to the signal and therefore we need to make assumptions on the noise properties. If the noise is assumed to be weakly stationary, Gaussian and ergodic with zero mean, it can be shown that the likelihood for parameters $\vb*{\lambda}$ is given by \cite{whittleAnalysisMultipleStationary1953}:
\begin{equation}
    \label{eq:likelihood}
p(s|h(\vb*{\lambda})) = p(s|\vb*{\lambda}) \propto \exponential \qty [-\frac{1}{2} \bra{s - h(\vb*{\lambda})}\ket{s- h(\vb*{\lambda})} ]\, ,
\end{equation}
where we have introduced the inner product $\bra{\cdot}\ket{\cdot}$
\begin{equation}
    \label{inner_product}
    \bra{a (t)}\ket{b (t)} =4 \Re \int _{0} ^\infty \frac{\tilde{a} ^* (f) \tilde{b} (f) }{S_n (f)} \, \dd f \, .
\end{equation}
The tilde indicates the Fourier transform and $S_n (f)$  is the one-sided spectral density of the instrumental noise, which can be interpreted as the size of the root mean square fluctuations at a given frequency $\Delta n _{\text{rms}} \sim \sqrt{S_n(f) \Delta f}$ measured in time intervals $\Delta t = 1 / \Delta f$ where $\Delta t$ is the observation time. 

In practice we do not know exactly what the signal $h$ is, so the main idea behind detection of a signal is to vary the parameters of our model $h(t;\vb*{\lambda})$ in order to find those which maximize the likelihood.

\paragraph{ The signal-to-noise ratio:} There are a variety of techniques used to detect signals depending on the type of source \cite{gairMakingSenseData}, but for the time being let us assume that a superior entity kindly provides us with the true waveform $h(t;\vb*{\lambda})$. Given a waveform $h$, we can calculate how loud the signal is with respect to the noise by calculating the signal-to-noise ratio $ \text{SNR} = \rho = \sqrt{\qty|\bra{h}\ket{h}|} $. The higher the signal-to-noise ratio, the easier it is to identify the signal in the data.\\

\paragraph{ The Fisher Information Matrix:} If we want to understand how accurately parameters can be determined, it is useful to calculate the Fisher Information Matrix defined as
\begin{align}
    \label{eq:fisher_matrix_def}
\Gamma ^{ij} &= \mathbb{E} \qty [ 
    \pdv{l}{\lambda_i} \, \pdv{l}{\lambda_j}
]= \bra{\partial_i h(t;\vb*{\lambda}_0)}\ket{\partial _j h(t;\vb*{\lambda}_0)}
\, ,
\end{align}
where $l$ is the log-likelihood, $\mathbb{E}$ denotes the expectation value of realisations of the noise, and the second equality follows for the likelihood given in eq.~(\ref{eq:likelihood}).
Formally, this provides a lower bound on the variance of any unbiased estimator $\hat{\lambda}$ of the parameters of the signal: $\text{cov}(\hat{\lambda}_i,\hat{\lambda}_j)\geq (\Gamma^{-1})_{ij}$.
In other words, given any unbiased estimator of the parameters, the variance of the estimators cannot be smaller than the elements of the inverse of the Fisher Matrix.
Therefore, The Fisher matrix provides a indication of how well parameters can be measured. For high SNR, the Fisher Matrix not only provides a lower limit, but gives a good approximation to the shape of the likelihood, which can be seen by expanding the likelihood for small deviations from the true parameters and keeping the dominant term.

From the form of the Fisher matrix we can deduce that those parameters that strongly influence the waveform, are those that might be more precisely determined. Furthermore, since the Fisher matrix scales as $\Gamma \sim \rho ^2$, the precision with which we can determine the parameters will be higher as the SNR increases, i.e., $\Delta \Lambda ^i \sim 1/\rho \rightarrow 0$ as $\rho \rightarrow \infty$.

\paragraph{ The posterior:} In a Bayesian framework, the parameters of the source are not regarded as fixed, but are considered as random variables themselves. This is one of the fundamental differences to the classical \emph{Frequentist} approach.  Therefore, given an observed gravitational wave strain $s$, it is possible to construct a distribution for the source parameters, the {\it posterior distribution} $p(\vb*{\lambda} |s)$,  using the Bayes theorem:
\begin{equation}
p(\vb*{\lambda} |s) = \frac{p(s|\vb*{\lambda}) \, p(\vb*{\lambda})} {p(s)} \, 
\end{equation}
where $p(s)$ is the {\it marginal likelihood}, or {\it evidence}, and $p(\vb*{\lambda})$ is the prior distribution which defines our prior knowledge on the parameters. 

A Bayesian posterior can be interpreted as a probability distribution that encodes our state of knowledge about the parameters of the model based on the observed data and prior information. To obtain the posterior distribution it is often necessary to construct a sequence of samples that has a distribution that follows the target distribution. Typically this is done using Markov chain Monte Carlo algorithms such as Gibbs sampling or the Metropolis-Hastings algorithm.
\cite{Ashton:2021anp,Meyer:2020ijd}

\paragraph{ Bayesian hypothesis testing} is a suitable framework for comparing different models, whose parameters are not known precisely and are therefore well represented as distributions which depend on the model. For a review, see e.g.~\cite{Thrane:2018qnx}.

If we have two models $M_1$ and $M_2$ that describe the relation between the parameters $ \vb*{\lambda}$ and the data $s$, we can compute the posterior odds ratio 
\begin{equation}
    O_{12} = \frac{ p(s | M_1 ) }{p(s | M_2 ) } \frac{ p(M_1 ) }{p( M_2 ) } 
    \qquad
    \qquad
    p(s| M_A ) = \int  p(s|\vb*{\lambda}_A, M_A ) \, p(\vb*{\lambda}_A\,| M_A) \, \dd \vb*{\lambda}_A
    \qquad \qquad A=1,2
\end{equation}
to understand which model is favored by a set $s$ of observations. The first ratio is called Bayes factor and only depends on the data and assumed model, whereas the second ratio is called the prior odds ratio and represents our prior belief on the correctness of each model. 

The Bayes factor is the ratio of the evidences of the two models, where each evidence $p(s | M_A )$ is the probability of observing the data $s$ under the assumptions of model $M_A$. The evidence is a quantity that combines the information given by the observed data and the prior belief on the parameters of a model. Given a set of observations, the likelihood  $p(s|\vb*{\lambda}_A, M_A )$ expresses which parameter regions are preferred by the data, whereas the prior $p(\vb*{\lambda}_A\,| M_A)$ represents the prior knowledge of the parameters according to the model $M_A$.

\paragraph{ Statistical and systematic errors}
The noise realization of a given observation leads to a statistical error, i.e., a difference in the maximum likelihood parameters, or peak of the posterior distribution, from the true parameters. Statistical errors are accounted for in the width of the posterior, scale with SNR like $\delta_{\text{stat} }\sim 1/\rho$ and would average out over hypothetical repetitions of the experiment. Systematic errors, on the other hand, arise from differences between the true data generating process and that assumed in the likelihood, e.g., from an approximation or error in the assumed waveform model. These are not accounted for in the posterior width, are independent of SNR and do not average out over repeated experiments. Systematic errors can lead to wrong conclusions for tests of GR and it is therefore important to use waveform models that keep them smaller than the statistical errors.

%%%%%%%%%%%%%%%%%%%%%%%%%%%%%%%%%%%%%%%%%%%%%%%%%%%%%%%%%%%%
\subsection*{Machine Learning Techniques}
%%%%%%%%%%%%%%%%%%%%%%%%%%%%%%%%%%%%%%%%%%%%%%%%%%%%%%%%%%%%
The central goal of gravitational-wave detector data analysis is to determine the system parameters. This is usually accomplished with Bayesian inference as previously described. 
One explores the parameter space and compares the waveform model to the observed strain by calculating the likelihood. 
Since this evaluation can be computationally expensive, there is a pressing need for new, fast and accurate Bayesian inference techniques. In the last years, the GW community has put a lot of effort into reducing the computational cost of the different stages of the analysis with machine learning techniques and exploiting graphics processing units (GPUs). Since there is a vast growing literature about these topics, in this section we aim at explaining some of the most promising techniques that have gained success in the last years.

It has been shown that gravitational waveforms exhibit redundancy in the parameter space. This suggests that the amount of information necessary to represent a fiducial waveform model is smaller than what might naively be expected, and the variety of waveforms in the parameter space can be accurately captured using remarkably few representative waveforms. These representative waveforms can be found by using different kinds of algorithms, and they comprise a reduced basis from which all other waveforms within the same physical model can be represented. The key step is to compute waveform projections onto the basis and obtain a simpler representation of the waveform model.
This allows a rapid online evaluation of the reduced model at the expense of having to build the basis prior to the application \cite{PhysRevX.4.031006, PhysRevLett.106.221102}. These techniques have been subsequently given rise to accurate surrogate waveform models \cite{Katz:2021yft, Chua:2020stf, Katz:2020hku, Chua:2018woh}. Other techniques go a step further and build reduced order quadrature models that allow computation of the inner product (Eq. \ref{inner_product}) at a much lower computational cost \cite{Canizares_2015}.

In \cite{Zackay:2018qdy}, it has been proposed a fast way of evaluating the likelihood. The key idea is that sampled waveforms with non-negligible posterior probability are all very similar to each other in the frequency domain, and differ only by smoothly varying perturbations. If we work directly with the ratios of the candidate and fiducial waveforms in the frequency domain, we can operate with a lower resolution without losing accuracy in computing the likelihood.

In the last few years, deep-learning techniques, such as likelihood free inference, have been shown to be particularly successful in directly sampling the parameters from the posterior \cite{Green_2020,dax2021realtime,green2020complete, Chua:2019wwt}.
The basic idea is to produce a large number of simulated data sets (with associated parameters), and use these to train a type of neural network to approximate the posterior. The trained network can then generate posterior samples extremely quickly once a detection is made. This bypasses the need to generate waveforms at inference time, thereby amortizing the expensive training costs over all future detections. A key difference, however, is the way in which the likelihood is used: for conventional methods, its density is evaluated, whereas in these techniques it is used to simulate data. This distinction is important when dealing with nonstationary or non-Gaussian detector noise, for which an analytic likelihood is either expensive or unavailable. In this case, one could nevertheless simulate data, in a noise-model-independent way, by injecting simulated signals into real noise.

%----------------------------------------------------------------------------------------------------
\section*{Future Challenges in gravitational wave data analysis}
%----------------------------------------------------------------------------------------------------

Over a little more than half a decade, we have learned a lot 
about analyzing the data of ground-based GW detectors. There 
have been advancements in a variety of topics related to GW data analysis, starting from waveform modelling, data handling, statistical modelling of the data themselves, and of course on parameter estimation and search algorithms. Future detectors will come with a different sensitivity curves than present days' observatories, and 
aim to measure GW signals in a lower frequency band. In the end, future observatories will present us with new challenges in the analysis of the data. We describe two of the most pressing here.

\subsection*{Overlapping sources, global fit and residuals and high SNR}

Further improvements to the sensitivity and bandwidth of ground-based detectors are already planned for the near future ~\cite{KAGRA:2020npa}, and preparations for next-generation detectors are underway, such as the European Einstein Telescope ~\cite{ET} and the Cosmic Explorer project in the US~\cite{CE}. These experimental developments will lead to a steep increase in the detection rate, and in addition, signals will linger for much longer times in the detectors’ sensitive band. It thus becomes likely that signals from different sources will be simultaneously present in the data. Recent estimates indicate that, for example, a typical binary neutron star signal will be overlapped by tens of binary black hole signals (e.g. ~\cite{Samajdar:2021egv}). Mock data analysis challenges have shown that it is still possible to detect overlapping signals, e.g. ~\cite{Meacher:2015rex}. However, studies of the resulting biases in the Bayesian parameter inference  ~\cite{Pizzati:2021gzd,Samajdar:2021egv} and computational challenges ~\cite{Smith:2021bqc} in this context have only recently begun.

A second notable example is the future LISA mission~\cite{LISA}. LISA will operate at frequencies between $1~\mathrm{mHz}$ and $0.1~\mathrm{Hz}$, and it is expected to observe a variety of types of sources. The first category are the supermassive black hole binaries (SBBH)~\cite{Klein:2015hvg} and the lighter stellar origin black holes (SOBHBs), which are currently visible to LIGO-Virgo. Then, there is the population of Ultra Compact Galactic Binaries (UGBs). These are almost monochromatic sources, mostly comprised by white dwarf binaries, and LISA is going to measure all of them simultaneously~\cite{Cornish:2003vj}. Finally, there is the extreme mass ratio inspirals (EMRIs) ~\cite{Babak:2017tow}, and possible stochastic GW signals originating from processes of the early Universe~\cite{Bartolo:2016ami,Caprini:2015zlo}. 

\begin{figure}[h]
\caption{{\it Left}: Summary of the potential sources measured by LISA, plotted on top of the detectors' sensitivity. Taken from~\cite{2017arXiv170200786A}. {\it Right}: A cartoon illustrating the {\it global fit} analysis scheme idea of the LISA data. Taken from~\cite{litten2020}.}
\label{fig:lisasens}
\centering
\includegraphics[width=0.55\textwidth]{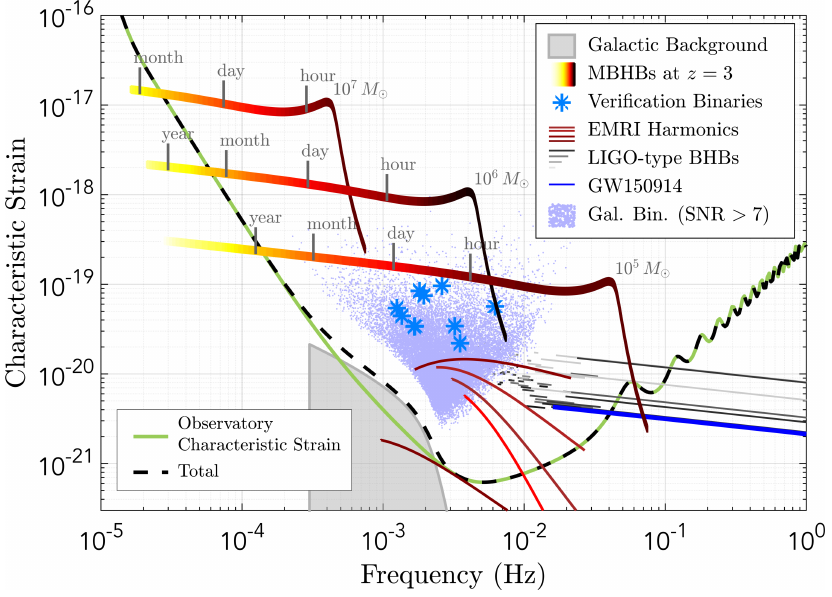}%GMS - Added path to actual graphics
\includegraphics[width=0.45\textwidth]{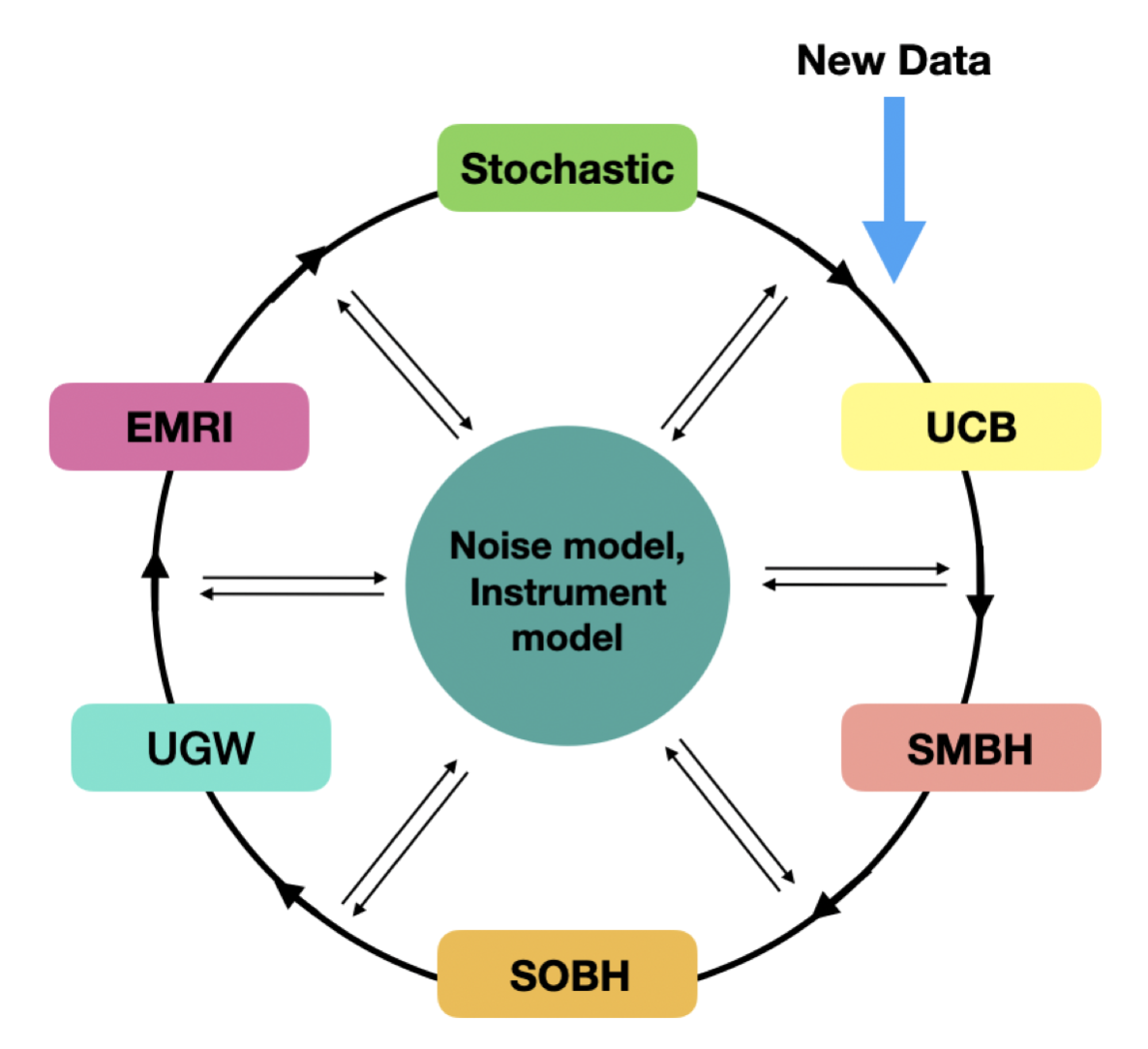}
\end{figure}

Taking the above into account, one immediately realizes that LISA is going to be a \emph{signal dominated observatory} (see figure~\ref{fig:lisasens}). Thus, one of the main data analysis challenges will be to identify and disentangle the GW signals, in the presence of possible non-stationary instrumental noise. For this purpose, the community has proposed a data analysis strategy based around a \emph{ global fit scheme} ~\cite{Cornish:2005qw}. The idea is represented in the cartwheel scheme on the right panel of figure~\ref{fig:lisasens}. As new data is transmitted to the ground, the first task would be to find and subtract the most abundant type of source, which are the compact binaries in the vicinity of our Galaxy. After this first step, another type of GW source will be targeted for analysis on the data residuals of the first step, after updating the information about the instrument noise model. In the end, all types of sources will be searched in the data, while more of them will be resolved as more data is available for analysis. 
The key idea is to refine the overall fit iteratively to reduce the number of undetected source.

This situation is ideal for trans-dimensional search algorithms, such as the reversible jump MCMC ~\cite{Green:1995}, where one can perform model selection and parameter estimation at the same time. This type of algorithms have already been demonstrated with LISA data, focusing mostly on the CGBs~\cite{Stroeer:2009hj}, but the complexity of the problem calls for new ideas~\cite{Teukolskyetal2019}. 

\subsection*{Non-stationary noise in LISA or 3G detector (glitches) and instrumental artifacts}

The effect of glitches, especially when coincident with an astrophysical signal, poses several challenges for gravitational-wave data analysis. Among the currently-used strategies is to subtract the artifacts of glitches from the data by modeling them with generic sine-Gaussian wavelets. To efficiently separate such glitches from the astrophysical signals requires a simultaneous modeling of both the glitches (using the wavelets) and the signal (using templates)
~\cite{Chatziioannou:2021ezd}. The contribution from the glitches determined in this way can then be subtracted from the data before passing it on for parameter estimation analysis. For more information about challenges and approaches for addressing nonstationarity in the noise, see e.g.
\cite{Edwards:2020tlp}.

\subsection*{Summary and outlook}
In conclusion, gravitational wave data analysis is a rich and rapidly evolving field. Many of the methods are interdisciplinary, yet require adaptations specific to the challenges of gravitational wave observations. As the experimental capabilities of the detectors improve over the next years, concurrent improvements in the data analysis strategies will be critical to robustly and efficiently respond to the data samples. This places significant demands on the data analysis strategies, requiring substantial advances over current methods. While progress is already underway, much more work remains to be done to handle longer signal durations, corresponding higher contamination by glitches, and overlapping signals. Moreover, the larger signal-to-noise ratios enabled by improved detector sensitivity requires a reduction in systematic uncertainties not only due to waveform models but also in the data analysis itself such as the sampling algorithms. In addition, to derive the greatest science impact from the measurements requires the algorithms to ultimately be capable of efficiently sampling yet higher-dimensional parameter spaces than currently possible, e.g. for eccentric, precessing, generic compact object binaries with higher modes, parameterized finite size effects and deviations from GR. Performing the analysis with diverse algorithms will also be essential to quantify biases. New approaches such as deep learning provide promising avenues for addressing many of the upcoming challenges in gravitational wave data analysis yet significant further work will be required to fully establish these methods as state-of-the-art for data analysis. 
%-------------------------------------------------------------------------------------------------
\printbibliography
\end{refsection}

\newpage
\begin{refsection}
\setcounter{figure}{0}
\setcounter{footnote}{0} 
\section{Ground-based Gravitational Wave Interferometers: Present and Future}
\begin{center}
    \textbf{Session chairs \& writers:}\\
    Archisman Ghosh\footnote{Ghent University, Proeftuinstraat 86, 9000 Gent, Belgium}\\
    Mikhail Korobko\footnote{Institut f\"ur Laserphysik \& Zentrum f\"ur Optische Quantentechnologien, Universit\"at Hamburg,
Luruper Chaussee 149, 22761 Hamburg, Germany}
\end{center}

\begin{center}
    \textbf{Invited speakers:}\\
    Laura Nuttall\footnote{University of Portsmouth, Portsmouth, PO1 3FX, UK}\\
    Bangalore Sathyaprakash\footnote{The Pennsylvania State University, University Park, PA, USA}
\end{center}

\subsection*{Overview}
Cosmology and astrophysics are undergoing a paradigm-shift, as more modern and precise instruments are bringing in new data. In the recent years black holes have been observed in gravitational waves (GWs)~\cite{abbott2016observation} and also directly in the electromagnetic spectrum~\cite{Akiyama2019}, in addition to probing their gravitational field~\cite{ghez1998high} with stellar orbits. These observations so far haven’t resolved many open questions about the properties of compact objects. The incompatibility of general relativity and quantum mechanics still prevents us from understanding the black hole information paradox or the nature of the black hole singularity. We do not know the origin of compact binaries and supermassive black holes. On the other hand, at cosmological scales, we still lack clues on the origin of dark matter, and any explanation of the accelerated expansion of the Universe. The precise value of the expansion rate itself remains elusive, since different measurement approaches yield statistically different results. This highlights the crisis in modern physics: despite the dramatic increase in quality and quantity of data, it has not yet brought the answers to these fundamental questions about the Universe.

In the past several years the new window to the Universe has opened, with the gravitational-wave observatories transitioning into the phase of regular observations. Currently, the catalogue of confirmed gravitational-wave events includes 52 mergers~\cite{abbott2019gwtc, abbott2021gwtc,abbott2021observation}, including mergers of binary black holes (BBH), binary neutron stars (BNS) and black-hole-neutron-star pairs. After the next instrumental upgrade in the O4 observing run of the Advanced LIGO-Virgo-KAGRA detector network the detection rates might reach as much as one event every 1-2 days~\cite{abbott2020prospects}. This new data might allow us to answer some of the fundamental questions mentioned above, contributing to resolution of the crises. The multi-messenger observations, where gravitational waves are observed alongside with EM and neutrino counterparts, allow us to test cosmology as well as physics of dense matter. Multi-messenger observations of neutron star merger GW170817~\cite{abbott2017gw170817} have already allowed scientists to exclude several models of dark matter~\cite{boran2018gw170817} and confirm the origin of heavy metals in kilonova explosions~\cite{abbott2017estimating}.

Despite these successes, the sensitivities of the current generation of detectors will not be sufficient to resolve many of the outstanding issues. Limited sensitivity to low-frequency signals may hinder early alerts for EM observatories. Limited sensitivity to high-frequency signals may prevent probing the final stages of NS mergers as well as post-merger oscillations, which are crucial for a study of the physics of dynamics of dense matter and an understanding of the equation of state governing the dense matter physics. Gravitational waves may allow us to resolve the Hubble tension using the standard siren approach~\cite{abbott2017Hubble}, but this requires significant statistics of hundreds of BBH events or tens of BNS mergers with EM counterparts~\cite{chen2018two}. 

Finally, there are sources of gravitational waves that are still to be observed. In particular, continuous gravitational waves from spinning neutron stars, GWs from supernova explosions and the stochastic gravitational-wave background.

The third generation of ground-based detectors (3G), e.g. Cosmic Explorer~\cite{reitze2019cosmic} and Einstein Telescope~\cite{punturo2010einstein}, will be more sensitive at all frequencies, and with increased sensitivity and detection rate new challenges will arise. Various open questions exist in our understanding of instruments, data analysis as well as general theoretical framework required to interpret the data. In the following, we provide a short overview of some of these questions.

\subsection*{Instrument}
3G detectors will feature sensitivity improvement by at least an order of magnitude across the broad frequency band. This will increase the observational reach of the detectors up to a redshift of $z\sim20$~\cite{bailes2021gravitational}.
The signals will be observed with very high signal-to-noise ratios (SNR), dozens high-SNR events a day. At the same time, the signals will remain in the detection band for much longer, hours or days, which would be enabled by the increased sensitivity at low frequency. 

This progress comes with several open questions. For example, there will be hundreds of overlapping signals in the detection band. Considering the complicated dynamical nature of the signals, it will come as a significant challenge for data analysis. Furthermore, the detector noise is currently considered to be Gaussian and “quasi-stationary” over a time scale of tens of seconds. This assumption is not necessarily true, and with much longer signals anticipated with the 3G detectors, non-stationary and non-Gaussian features in the detector noise, including detector “glitches”, will play a significant role. Going around this hurdle would require a paradigm shift in detector characterization and also the analysis methods including searches and source properties estimation. Along with this, the calibration of the detector would become significantly more complicated. In order to account for the non-stationary behavior of the detector, it will become necessary to acquire the calibrated state over days and weeks during which the signal remains in the detection band.

The most important technological challenge in reaching the desired sensitivities will become lowering the noise sources at low frequencies. This relies on the development of several key technologies, which currently have not reached the required quality (primarily mirror coatings~\cite{granata2020progress}, suspensions~\cite{nawrodt2011challenges}, quantum noise suppression~\cite{Danilishin2012,Danilishin2019}, seismic noise mitigation~\cite{harms2015terrestrial} and control systems). On the other hand 3G facilities are planned as the overall infrastructure for the future upgrades of the detectors for the next 50 years. 

Even at the facility limit of 3G detectors, not all possible phenomena will be explored. Some, like supernovae or stochastic sources, may require significantly higher sensitivities. Despite that, 4G detectors are not currently foreseen, with decades of the lifetime for the 3G network. At the same time, different approaches to GW detection emerge, based on smaller, cheaper detectors, potentially capable of observing GWs at higher frequencies. The field of applied GW astronomy is just at its infancy, and future years will undoubtedly bring many technological breakthroughs for small-scale detectors.

\subsection*{Data analysis and algorithms}
A number of questions and challenges emerge purely from the data-analysis side following the planned upgrades to the detectors and/or in the proposed 3G detectors. In this section and the following, we list a number of items that came up for discussion during the workshop and/or otherwise important for the community to understand.
\begin{itemize}
    \item \textit{Too much data!} Compact binary signals have become more and more frequent as the LIGO/Virgo detectors have progressed from O1 to O3 sensitivities. With O4 and A+ sensitivities we surely expect many more signals -- with A+ sensitivity, we expect at least 5 times the current number of compact binary events. This would be beyond what can be efficiently handled by current data analysis and data handling paradigms. Already in O3, we have seen cases where researchers external to the LVK have come up with interesting ideas beyond what was put out by the LVK. Would making data open solve the problem? Perhaps an idea would be to put out catalogues with limited science results and data and leave the community to further analyze and interpret them.
    \item \textit{Overlapping signals}. As already mentioned, multiple signals will be present in the detector at the same time.  How would the search and source properties estimation pipelines need to be adapted to this? How would this affect the prospects of other observations, e.g, supernova bursts, continuous GWs, or the stochastic background?
    \item \textit{The null stream}. The proposed triangular configuration of the ET would provide a “null stream” which would not contain any GW signals. Something that shows up in the null stream is therefore not an astrophysical signal, but a detector noise artefact. How can we best make use of the null stream?
\end{itemize}

\subsubsection*{Open questions in data analysis and algorithms, which came up during the workshop:}
\begin{itemize}
    \item Are current algorithms for detection and parameter estimation suitable for the signal-rich data from 3G observatories? What new paradigms in data analysis do we need to efficiently detect and characterize signals in 3G detectors?
    \item How do signal-rich data hinder the detection of signals from rare events such as SN bursts or pulsar glitches? Do they hinder the detection of continuous wave (CW) signals?
    \item How does the population of overlapping signals affect parameter estimation, discriminating astrophysical models, testing general relativity (GR), measuring cosmological parameters, etc.?
\end{itemize}

\subsection*{Observations and theory}
With increased sensitivities not only would we have more observations, but we also need to be prepared for new discoveries. Among challenges, there are many outstanding problems that could already be seen in the current detections. In particular, the coming years might bring us better understanding of the population of stellar-mass compact binary mergers: BBH mass spectrum, BBH formation channels and the change of merger rate with redshift. 

There is currently much uncertainty in our understanding of neutron star dynamics. We do not know what happens after the merger, whether phase transitions from hadrons to deconfined quarks occur and how the core behaves. Many of the analyses are very model-dependent. With the increased event rate and detection precision in 3G detectors, it will become crucial to choose the models or refine the criteria for such choice.

\begin{itemize}
    \item What might be the next big discovery? It could be the stochastic GW background; probably not primordial GWs, but the astrophysical background.
    \item For NSs, multimessenger as well complementary experiments (including, e.g., NICER) would give us a wealth of physics (see, e,g,~\cite{dietrich2020multimessenger}).
    \item What are the prospects for detecting continuous GW sources, e.g., from asymmetric NSs? Perhaps we would detect objects beyond the known pulsars. 
\end{itemize}

\subsubsection*{Open questions in theory, which came up during the workshop:}
 \begin{itemize}
     \item Why is the spin of the remnant of an equal-mass Schwarzschild merger ~0.7? The standard answer is that this is equal to the orbital angular momentum that is absorbed by the remnant. In any case, NR predicts it to be so! However it is possible that there is a deeper theoretical reason why this is so. It could be related to the entropy of a black hole or with what happens when two extremal Kerr black holes merge.
     \item How does the dynamical horizon that results from the merger of two black holes settle down to its final Kerr horizon? Can it be described by the full spectrum of quasi-normal modes (QNM)? What is the relation between the excited quasi-normal modes and the parameters of the progenitor binary? More broadly, what is the relation between the perturbation and the amplitude of the QNM spectra that are excited? Is there a uniqueness theorem here, i.e. one-to-one mapping from perturbation to QNM spectra?
    \item What are the signatures of phase transition from hadronic to free quarks in neutron star (NS) cores? What biases are introduced by quasi-universal relations in determining the NS equation-of-state?
    \item Many different tests of GR have been proposed. Are they testing different aspects of GR or do they have a common ground? What is the most effective way to test GR?
    \item What are the multipole moments of a dynamical horizon formed from the merger of a pair of Kerr black holes? How are they related to the observed signal at spatial infinity? 
 \end{itemize}
\subsection*{Conclusion and outlook}
The third generation of gravitational-wave detectors will bring many exciting scientific challenges. Tackling some of them will require a shift in our paradigms. These shifts need to be prepared for, starting already now. Early career scientists will play the crucial role here; driving the innovation and building up critical knowledge. There are a number of open open problems, in instrumental science, data analysis and theory. Solving them will require reaching across the different collaborations, developing the new methods and working in cooperation at multiple frequency bands of different GW detectors: space-borne, ground-based and PTAs.
\printbibliography
\end{refsection}

\newpage
\begin{refsection}
\setcounter{figure}{0}
\setcounter{footnote}{0} 
\section{Space-based Gravitational Wave Interferometers: LISA}
\begin{center}
    \textbf{Session chairs \& writers:}\\
    Jean-Baptiste Bayle\footnote{Jet Propulsion Laboratory, California Institute of Technology, Pasadena, CA 91125, United States}\\
    Daniela Doneva\footnote{Theoretical Astrophysics, Eberhard Karls University of Tübingen, Tübingen 72076, Germany}\textsuperscript{,}\footnote{INRNE - Bulgarian Academy of Sciences, 1784 Sofia, Bulgaria}\\
    Valeriya Korol\footnote{Institute for Gravitational Wave Astronomy \& School of Physics and Astronomy, University of Birmingham, Birmingham, B15 2TT, UK}\\
    Sweta Shah\footnote{Max-Planck-Institut für Gravitationsphysik (Albert-Einstein-Institut), D-30167 Hannover, Germany}\\
    Martina Muratore\footnote{Dipartimento di Fisica, Università di Trento and Trento Institute for Fundamental Physics and Application/INFN, 38123 Povo, Trento, Italy}\\
\end{center}

\begin{center}
    \textbf{Invited speakers:}\\
    Antoine Petiteau\footnote{Université de Paris, CNRS, Astroparticule et Cosmologie, F-75006 Paris, France} \\
    Elena Maria Rossi\footnote{Leiden Observatory, Leiden University, PO Box 9513, NL-2300 RA Leiden, the Netherlands} \\
    Chiara Caprini\footnote{Université de Paris, CNRS, Astroparticule et Cosmologie, F-75006 Paris, France}
\end{center}

This session was dedicated to the future space-borne \gls{gw} detector \gls{lisa}, including a description of the instrument, the various astrophysical and cosmological sources we expect in the \gls{lisa} frequency band, as well as the associated challenges in data analysis. The first three sections are summaries of the talks given respectively by Antoine Petiteau, Elena Maria Rossi, and Chiara Caprini. Then, we give the transcripts of a selected set of questions and answers that emerged during the following Q\&A sessions.

\subsection*{LISA mission instrument}

In this talk, Antoine gave a description of the \gls{lisa} instrument and its main measurements. He then presented a brief overview of the post-processing algorithms and the main challenges associated, opening a discussion on the links between instrumental effects and data analysis.

The goal of the \gls{lisa} mission is to to measure \glspl{gw} in the \si{\milli\hertz} frequency range. A wide variety of astrophysical and cosmological sources are expected to emit in this frequency band (c.f. the following sessions for more information on these sources and the associated detection methods). The measurement of \glspl{gw} is achieved by tracking the relative distance between the pairs of 6 free-falling \glspl{tm} at the \si{\pico\meter} precision. \Gls{lisa} is designed to be a constellation of three identical spacecraft in a triangular formation, trailing the Earth on a heliocentric orbit. Six active laser links connect the spacecraft over \SI{2.5}{Mkm} armlengths. The distance between the \glspl{tm} is monitored by continuously operating heterodyne laser \glspl{ifo} in both directions along each inter-satellite link.

The \glspl{tm} are shielded by their hosting \gls{sc} to protect them from spurious forces that mimic the effect of \glspl{gw}, and therefore reduce the instrument sensitivity. Because the \glspl{tm} must remain in free fall, the \glspl{sc} continuously adjust their positions using \si{\micro\newton}-thrusters: if a \gls{sc} moves away from the \glspl{tm}, its relative motion is detected and the \gls{sc} is steered. This technology has been tested on flight by \gls{lpf}. \Gls{lpf} had two readouts of the \gls{tm} position: interferometric and capacitive sensing. \Gls{lpf} demonstrated the noise level that we require for \gls{lisa}. In \gls{lpf}, the high-frequency sensitivity is limited by the \gls{ifo} noise, while the mid-frequency sensitivity is limited by molecules hitting the \gls{tm}. Other noises, e.g., due to temperature, are not yet fully understood. 
 
To fulfil the scientific program of the mission, each spacecraft hosts mainly an on-board computer and phasemeter, two laser sources, and two identical assemblies of roughly cylindrical shape, which are called \glspl{mosa}. They are mounted in a common frame that allows rotation of each assembly about the vertical axis in order to accommodate the breathing of the constellation induced by orbital dynamics, i.e., to adjust the pointing direction to the distant spacecraft, from the predetermined \SI{60}{\degree} to $\pm \SI{1}{\degree}$. Each \gls{mosa} contains a telescope, an optical bench, and a \gls{grs} enclosing a \gls{tm}. Laser sources produce \SI{2}{\watt} laser beams, which are then sent to the distant \glspl{mosa} by the telescope. Because of beam divergence over the \SI{2.5}{Mkm} arms, the distant telescope collects around \SI{100}{\pico\watt} of light and generate a new laser beam, which is then redirected to the optical bench. Therefore, contrary to ground-based interferometer, \gls{lisa} cannot use passive reflection for the counter-propagating beam; instead, each spacecraft operates like an active transponder: a new high-power beam is generated, phase-locked to the incoming weak beam with a fixed offset frequency, and sent back. One laser source in the constellation is chosen as the primary laser, and its frequency is stabilised to a reference cavity. All other lasers are phase-locked to it with a set of frequency offsets, given by a pre-computed frequency plan \cite{geh}.

To track the relative distance between the pairs of \glspl{tm} \Gls{lisa} uses the split interferometric measurement system: the distance from one \gls{tm} to another is decomposed into three measurements. To this end, each optical bench hosts three \glspl{ifo}: the inter-spacecraft (also called long-arm, or science) \gls{ifo} measures the distance between two spacecraft with the interference between the distant and the local laser beams. The test-mass \gls{ifo} measures the distance between the spacecraft and the \gls{tm}, as it is the interference between the two adjacent lasers, one bouncing on the local \gls{tm}. Lastly, the reference \gls{ifo} is used in post-processing as the interference between two adjacent lasers, without any \gls{tm} motion tracking. These are 18 interferometric measurements in total. In addition to these measurements, we estimate the inter-spacecraft distance (called ranging) by modulating the laser beams exchanged by the \glspl{sc} with a pseudo-random noise code, which we can then correlate. We also modulate the beams (creating sidebands) using the clock jitter in order to correct for clock noise in post-processing.

Various instrumental noises (and artifact) enter those measurements. Laser stability noise is the dominant source of noise. Ground-based detectors have equal and constant armlengths, such that this laser noise cancels when the two beams are recombined. \Gls{lisa} has unequal and time-varying armlengths (due to orbital mechanics), and laser noise expected at 8 to 9 orders of magnitude above \gls{gw} signals. Therefore, one key element in the \gls{lisa} data production chain is the post-processing technique dubbed \gls{tdi}. \Gls{tdi} time-shifts and combines various measurements in order to reduce laser noise to required levels, circumventing the impossibility of physically building an equal-arm interferometer in space.

Other noises can be measured independently and then suppress in post-processing, in the initial noise reduction pipeline, while others will stay in the data and limit the sensitivity. Important questions and efforts are dedicated to the on-the-fly estimation of the noise (as they mostly cannot be measured on ground in a satisfactory way) and how to mitigate instrumental artifacts, such as gaps or glitches.

\subsection*{Astrophysics with LISA}

In the talk, particular attention was paid to \glspl{smbhb}, \glspl{wdb}, and \glspl{emri}.
 
\Glspl{smbhb} (especially with a mass between \num{E4} and \num{E7}) are regarded as the loudest sources to be seen with \gls{lisa}. They are expected to stay in the \gls{lisa} band for weeks to months depending on the properties of the binary system, such as its mass. On galactic scales the \glspl{smbhb} will tell us about the formation and evolution of galaxies. The detection rate is not well constrained – it ranges between 10 and 300 events per year. An exciting possibility for high precision multi-messenger astrophysics can be realized through a joint Athena-\gls{lisa} detection for certain types of sources. Athena is a space-based X-ray observatory expected to be launched around 2030 and thus the two missions are planned to be jointly operating for some time. 
 
Another exciting class of sources detectable with \gls{lisa} are the \glspl{emri} that constitute a stellar-mass object, such as black hole or brown dwarf, orbiting around a \gls{smbh}. They can be viewed as test particles probing the innermost and highly curved region around \gls{smbh}. Such sources will stay in the \gls{lisa} band for years and the detection rate is estimated to be a few to \num{E3} events per year. Through \glspl{emri}, we can also study the content of the galactic nuclei. For instance, the rate at which these events occur depends on the existence and characteristics of a compact cusp of a compact object orbiting around a \gls{smbh}. It is also believed that the formation of \glspl{emri} is assisted by the presence of gas around the \gls{smbh}.
 
Solar mass compact objects binaries are the most numerous sources in the \gls{lisa} band. These are persistent sources staying in the \gls{lisa} band for the whole duration of the mission. The higher frequency sources (higher than say $10^{-3}$ Hz) will be likely resolved. These include \glspl{wdb} with periods less than 15-20 minutes. The low-frequency binaries will form a background. It is estimated that \gls{lisa} will observe tens of thousands of \glspl{wdb}. As a matter of fact, we have already identified through electromagnetic observations several such \glspl{wdb} that will be used for instrument calibration of \gls{lisa}. Through the observations of such objects, we can learn more about their tidal interaction, the stellar binary evolution, accretion discs around such objects, population studies, star population history, the galactic structure, etc. With \gls{gw} observations, we can also peer through dust regions of our Milky Way Galaxy prevented by standard \gls{em} observations.   
 
\gls{lisa} will also be able to detect planets orbiting around white dwarf binaries. What will be actually observed is a periodic Doppler modulation of the \gls{gw} signal. This frequency shift will be measurable for Jupiter-like planets at a distance smaller than \SI{10}{\astronomicalunit}. Similarly, brown dwarfs orbiting around \glspl{wdb} can be detected that can shed light on the missing link between planets and stars. Brown dwarfs will be detected more abundantly than planets by a factor of 30 to 150.

\subsection*{Cosmology}

\Gls{sgwb}, potentially detectable by \gls{lisa}, can be produced by high energy phenomena occurring in the early universe. \Glspl{gw} can also probe the accelerated expansion of the late universe through observations of merging binaries and \glspl{emri}. 

Possible \gls{gw} sources acting in the early universe lead to the production of \gls{sgwb} whose detection will allow us to probe the big bang at extremely early times, between inflation and \gls{bbn}, that are otherwise inaccessible through electromagnetic radiation. We are actually not sure that a \gls{lisa} detectable \gls{gw} signal exists, due to the fact that the source models rely on new untested physical scenarios. There are a number of proposed models, though, that predict such SGWB.

Another intriguing possibility is to set constraints on primordial black holes (PBH) with \gls{lisa}. These objects, within a certain mass range, are still a viable dark matter alternative. It turns out that second-order scalar field perturbations giving rise to PBH in the relevant mass range, produce an observable SGWB detectable by \gls{lisa} and thus the hypothesis for their existence can be tested.
 
Phase transitions related to the grand unification might leave a network of cosmic string producing \gls{gw}. The \gls{sgwb} from such cosmic strings is within the \gls{lisa} sensitivity band. On the other hand, LISA can also provide tests of scenarios beyond the standard model of particle physics complementary to particle colliders.

From all this it is evident that  even though \glspl{sgwb} from the primordial universe  might seem speculative, their potential to probe fundamental physics is great and amazing discoveries can be around the corner.

\Glspl{gw} can be also used to probe the late time expansion of the universe and determine the Hubble constant. The \glspl{gw} are analogous to standard candles in astrophysics and can test the expansion of the universe. An advantage compared to the standard candles is that the measurement of the luminosity distance is easy and direct. Measuring the redshift directly from the \gls{gw} observations, though, is just not possible due to the mass and redshift a degeneracy in the determination of the parameters of the \gls{gw} signal of merging compact objects. We can use alternative methods, such as identification of the host galaxy via a measurement of the transient electromagnetic counterpart or a statistical method where a cross correlation of the sky position given by the \gls{gw} measurement with galaxy catalogues is performed. \Gls{lisa} sources with expected EM counterpart where the former approach is applicable are massive black hole binaries at the center of galaxies while the statistical method can be used with \glspl{emri}.

\subsection*{Q\&A session: LISA mission instrument}

\textbf{Q:} What is the designed pressure and temperature inside the test masses chambers? Could sublimation occur? \\
\textbf{A:} The vacuum inside the test-mass changer is not as good as the vacuum in outer space. Thus, it will be vented to space so the pressure will decrease along the mission. The temperature is expected to be quite stable. The Brownian noise will decrease slowly until a plateau is reached and in any case, it is easy to model.

\mbox{ } \\
\textbf{Q:} Could imperfections in the vacuum between the satellites be a source of noise? How could this be corrected in the data analysis? \\
\textbf{A:} Some of the ionized particles can change the optical index between spacecraft and contribute to the noise. We expected it to be one order of magnitude below the requirements. This will be studied more.

\mbox{ } \\
\textbf{Q:} What will be the data stream rates transmitted from \gls{lisa}? Will there be many auxiliary channels? Will data be stored on board? 
\textbf{A:} Most of the data is telemetered to Earth (after it is downsampled onboard to \SI{4}{\hertz}). The science data comes first, then housekeeping data. About 8 hours a day streams of data, with an online analysis (fast) as we get the data. So the time between the data containing an event is downloaded to a first alert would be 10s to minutes. Worst case, the event happens after streaming, so you add 14 hours till the next transmission.

\mbox{ } \\
\textbf{Q:} What are the requirements on the measurement precision of the angle between the laser beams in \gls{lisa}? Can this error affect the localization of sources?
\textbf{A:} We are measuring these angles, and the telescopes sending the beams are locked to point correctly to the distant spacecraft. The main issue is the jitter of the spacecraft, which must be suppressed as part of \gls{tdi}.

\mbox{ } \\
\textbf{Q:} How much time will pass between \gls{lisa} seeing a signal and us knowing about it? Including getting the data, applying the \gls{tdi} algorithm, etc.? 
\textbf{A:} For the high \gls{snr} sources, such as the \glspl{smbhb}, it will take between a few tens of minutes for the optimal case and up to 16 hours in the worst case. In the case of low \gls{snr} sources, it can naturally take longer because we will need to accumulate data and dig inside these data.

\mbox{ } \\
\textbf{Q:} Are the systems fully duplicated? How is electronics protected from cosmic rays and other sources of degradation/errors? Is the performance expected to reduce over time?
\textbf{A:} There’s a high level of redundancy, e.g. duplication in electronics, laser, thrusters, star trackers and so on. Systems are also space-qualified, so resistant to cosmic rays. We will have degradation, therefore a noise that changes with time (slightly reduced for most systems), which is not yet accounted for in the science analysis we have.

\mbox{ } \\
\textbf{Q:} What is the \gls{lisa} mission duration? How do we handle gaps during the observations (that is a problem appearing due to the fact that we aim to observe sources with very long signals)? 
\textbf{A:} We expect roughly 4 years of usable data, but 10-25\% of the time there will be no data collected. However we will introduce protected periods -- when you know that something interesting will happen the maintenance will be shifted.
The total duration of the mission with extension will be up to 10 years.

\mbox{ } \\
\textbf{Q:} From the discussions, one is left with the impression that the complexity of \gls{lisa} data is different compared to the ground-based ones? Is this true since ground detectors can also have a very large number of events? 
\textbf{A:} This applies only to LIGO and VIRGO. For 3G detectors, the situation will be similar to \gls{lisa} with \SI{1000}{\second} of overlapping sources

\mbox{ } \\
\textbf{Q:} How much data will \gls{lisa} generate? How does that compare to LIGO-Virgo? And say the Event Horizon Telescope? 
\textbf{A:} For raw L0 data we expect 1GB per day. This is mainly limited by the band of the instrument. After \gls{tdi} - a few GB per day that after further analysis will become TB or even PB of data in total for the \gls{lisa} mission. For some \gls{gw} sources the data should be accumulated for a long time and then analyzed, which is a challenge. We expect similar challenges only from ET, but not LIGO.

\mbox{ } \\
\textbf{Q:} Can \gls{lisa} techniques be used for 3G detectors?
\textbf{A:} We are trying to connect with ground-based detectors and collaborate. 

\mbox{ } \\
\textbf{Q:} In practice how would one estimate the noise \gls{psd} of \gls{lisa} if we have a source dominated signal? 
\textbf{A:}  There are a number of ways being studied by \gls{lisa} Data Processing and \gls{lisa} Science \gls{tdi}. One of the ways to estimate \gls{lisa} noise might be through \gls{tdi} channels that are more sensitive to noise and less to signal. There are a certain combinations of TDI channels which is least sensitive \gls{gw} signal and most sensitive to instrumental noise, and this is one of the known ideas for estimating \gls{lisa} \gls{psd}.

\subsection*{Q\&A session: Astrophysics with LISA, waveform modeling, and data analysis}
\textbf{Q:} To accomplish the science objectives for astrophysics, how well do we need to model \gls{gw} signals? Source confusion or inaccurate waveform modeling can cause biases in the recovered parameters. How small should these biases be in order to use \gls{lisa} observations for astrophysics studies?
\textbf{A:} Of course an error in the model or calibration will cause biases in the parameters. For astrophysicists, the precision required makes these errors negligible, except for sky localization and distances. Ongoing work in the consortium is performed to exactly answer these questions.

\mbox{ } \\
\textbf{Q:} How reliable are the \gls{smbhb} population studies and are they compared with galaxy catalogs? 
\textbf{A:} Black hole populations are well constrained against electromagnetic observations, but only up to roughly redshift $z\sim 5$. This means that synthetic catalogs of \gls{gw} sources are highly uncertain beyond redshift 6 or so.

\mbox{ } \\
\textbf{Q:} For \glspl{smbhb} with, say, $10^4<M<10^7$, is \gls{lisa} going to have selection biases (inclination, sky location, mass ratio, spins, eccentricity, etc.)? Or are the \glspl{snr} going to be so large that if \glspl{smbhb} are there, we'll see them for sure?
\textbf{A:} \Gls{lisa} is not sensitive the same way depending on the localization of the source. For example, we are ``blind’’ if the source is exactly in the plane of \gls{lisa}. Inclination, orientation, etc. also impact the \glspl{snr} by a factor of roughly 2 or 3.

\mbox{ } \\
\textbf{Q:} With so many different sources with complicated dynamics (like extreme mass ratio waveforms) with the rates of thousands per year, how will different events be separated in data analysis? 
\textbf{A:} High \gls{snr} sources can be distinguished, but the rest will form a background. For \glspl{emri}, we’re currently trying to fit with phenomenological models which are not optimal.

\mbox{ } \\
\textbf{Q:} How many stellar mass black-hole inspirals do you still expect to see with \gls{lisa}?
\textbf{A:} In 2016, we expected up to 100 binaries detected in \gls{lisa} that will merge in LIGO. Now, we lost a factor of 4 because of the integrated sensitivity at high frequency and another factor of 3 because of reduced rates in the population based on the recent data from LIGO and VIRGO. So now we expect a few to several.

\mbox{ } \\
\textbf{Q:} How many modes can we observe for different sources? For \glspl{emri}, we surely need many. How about the other sources? 
\textbf{A:} For galactic binaries we do not expect that we can measure any higher modes. For \glspl{smbhb}, there are sources with a particular configuration of high mass ratio and \gls{snr} for which such measurement is possible, but there are no dedicated studies on this problem. Indeed, there is work to do there. In the \gls{ldc}, we are planning to use 100 modes for \gls{emri}, for \gls{smbhb} we use 8, especially for very massive ones. This is because higher modes are in the central part of the \gls{lisa} sensitivity band. Higher modes are starting to be considered as well in cosmology applications. For example, we are also extending our studies of standard sirens to higher modes.

\mbox{ } \\
\textbf{Q:} How to tackle the global fit problem? 
\textbf{A:} Not many people are looking at the global fit problem and this should be done; in the next \gls{ldc}, we are introducing various sources (galactic binaries and \glspl{smbhb}).

\mbox{ } \\
\textbf{Q:} How far away would we be able to observe planets? Which will be the frequency band? 
\textbf{A:} \Gls{lisa} will be able to observe planets everywhere in the Milky Way, and much further than electromagnetic observations which basically can detect planets only in the Solar neighbourhood. Note however that \gls{lisa} will detect only the \gls{gw} signal emitted by white dwarf binaries ($\sim \SI{E-3}{\hertz}$), and will infer the presence of a planet through the Doppler perturbations in the \gls{gw} signal given by the circumbinary planets. See https://arxiv.org/abs/1910.05414 for more details.

\mbox{ } \\
\textbf{Q:} Can we use population studies of \glspl{dwd} to infer properties of the Milky Way dark matter halo? 
\textbf{A:} Not quite the dark matter halo, but unresolved and also resolved \glspl{wdb} can be used to constrain the baryonic distribution of matter (see e.g. https://arxiv.org/abs/1912.02200).

\mbox{ } \\
\textbf{Q:} Are eccentric binaries included in the waveform modeling? 
\textbf{A:} At the moment no, but this is important work under development.

\mbox{ } \\
\textbf{Q:} What source parameters \gls{lisa} can and cannot measure compared to e.g. LIGO/Virgo? 
\textbf{A:} For \glspl{bhb}, we span a very different mass range compared to LIGO/Virgo. The redshift distribution is also very different. Neutron-stars binaries and stellar-mass binaries in the inspiral phase can be also detected by \gls{lisa}, but should be massive enough and local. Up to a few tens of objects will be detected for the whole mission. So practically we can measure the same parameters but for different sources. 

\subsection*{Q\&A session: Cosmology} 

\textbf{Q:} How do you plan to look for these different stochastic sources taking also into account the uncertainty we might have on the noise? Are you going to look for each model at time with a sort of match filtering and see when the model matches with the data? Are there other data analysis techniques you plan to use?
\textbf{A:} This is an absolutely fundamental question that has not been yet fully addressed even though there are attempts. It is not clear whether searching for the signal one by one using matched filtering is doable because of the uncertainties in the predictions of the primordial signals. So we do not have clear predictions about the shape and the amplitude of the signals. The approach we have been following by now is to make an agnostic identification of the \gls{sgwb} proceeding bin per bin. This means that one can divide the frequency band of \gls{lisa} into several bins. In each bin, one first has to clear the astrophysical signal, and fit the remaining with a model for the \gls{sgwb} and the noise, assuming that the stochastic signal is a simple power law. Then we have only two parameters for the stochastic signal to fit that are the amplitude and the spectral index. At the same time, we will need a noise model, that is something yet unclear. By repeating this operation bin by bin, we can theoretically reconstruct the whole \gls{sgwb}.

\mbox{ } \\
\textbf{Q:} If \gls{lisa} would not observe any background signal from \glspl{pbh}, would this non-detection rule out the existence of these objects? 
\textbf{A:} \Gls{lisa} is only sensitive to a given mass range, so it can only put constraints on this range. Also, the production of \glspl{pbh} relies on some assumptions about the primordial curvature power spectrum. By changing this ansatz, the predictions change as well. Thus \gls{lisa} will provide constraints in a certain mass range of \glspl{pbh}, but certainly can not rule them out completely. 

\mbox{ } \\
\textbf{Q:} There are several different sources of stochastic background, with similar spectral shapes (i.e. broad, no sharp features). How do we tell one source from another when we see it in the data?
\textbf{A:} Typically the different \gls{sgwb} sources will create signals with different spectral indices, nominally allowing us to differentiate between models. That said they'll be overlapping in frequency space so the question of how to simultaneously analyze 2+ backgrounds if multiple are present is an ongoing problem.

\mbox{ } \\
\textbf{Q:} If we have multiple \glspl{sgwb} overlapping, how are we going to disentangle them? 
\textbf{A:} There is no clear way to distinguish/subtract the astrophysical and cosmological type of backgrounds. We expect different power laws and spectral shape but recent studies show that this might not be the case and overlapping is possible.

\subsection*{Conclusion}

\Gls{lisa} is a unique gravitational wave observatory that is expected to fly roughly in 10 years from now. It will be sensitive to a completely different gravitational wave frequency range compared to the current and future planed ground based detectors, and to PTAs. Thus, LISA will open a window for investigating a large variety of new astrophysical and cosmological gravitational wave sources covering all redshifts: from the early-Universe to the present-day-Universe. Therefore, \Gls{lisa} is offering a huge scientific potential to test various unresolved questions in astrophysics, cosmology, and fundamental physics. A large group of scientists is now working to resolve the open problems and challenges that the mission poses, and is currently on track to see this astonishing observatory operating soon and challenging our understanding of the Universe.

%\printbibliography
\end{refsection}

\newpage
\begin{refsection}
\setcounter{figure}{0}
\setcounter{footnote}{0} 
\section{Pulsar Timing Arrays}
\begin{center}
    \textbf{Session chairs \& writers:}\\
    Golam Shaifullah\footnote{Dipartimento di Fisica ``G. Occhialini'', Università di Milano-Bicocca, Piazza della Scienza 3, 20126, Milano, Italy}\\
    Lijing Shao\footnote{Kavli Institute for Astronomy and Astrophysics, Peking University, Beijing 100871, China}\textsuperscript{,}\footnote{National Astronomical Observatories, Chinese Academy of Sciences, Beijing 100012, China}
\end{center}

\begin{center}
    \textbf{Invited speakers:}\\
    Caterina Tiburzi\footnote{ASTRON - the Netherlands Institute for Radio Astronomy, Oude Hoogeveensedijk 4, 7991 PD, Dwingeloo, The Netherlands}\\
    Michael Kramer\footnote{Max-Planck-Institut für Radioastronomie, Auf dem Hügel 69, D-53121 Bonn, Germany}\textsuperscript{,}\footnote{Jodrell Bank Centre for Astrophysics, University of Manchester, Manchester M13 9PL, UK}
\end{center}

\begin{figure}[ht!]
    \centering
    \includegraphics[width=0.98\textwidth]{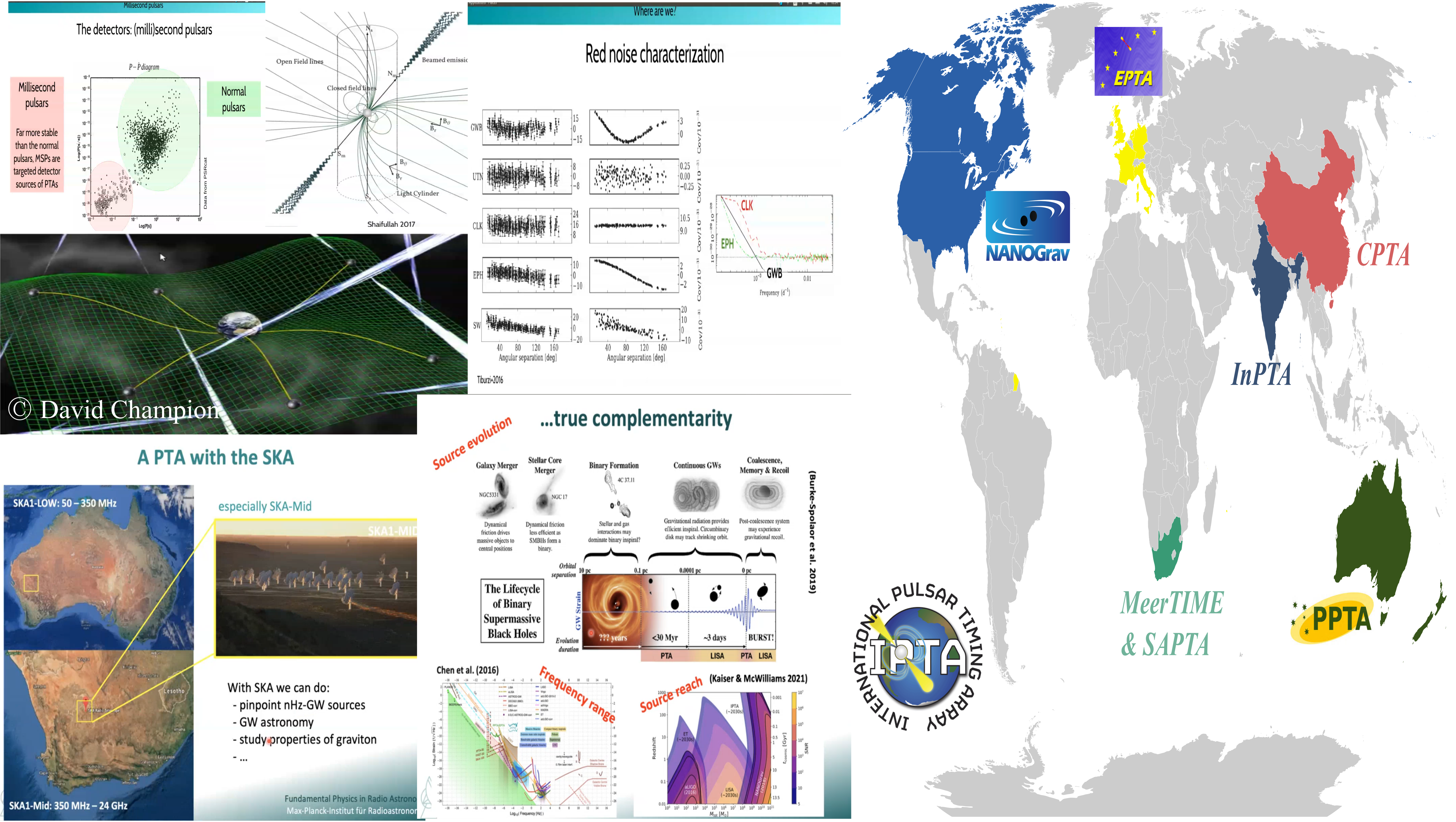}
    \caption{Left - Montage from the slides presented, highlighting the nature, scope, dominant noise sources and future of GW science with PTAs. Right - The International PTA collaboration and their principal, constituent members. }
    \label{fig:my_label}
\end{figure}

\subsection*{Executive Summary}
The pulsar timing array (PTA) session had two talks presented by Dr. Caterina Tiburzi, a post-doc at ASTRON, who has been looking at mitigating the effects
of the Solar Wind on PTA data and Prof. Michael Kramer, one of the directors
of the Max Planck Institute for Radio astronomy in Bonn, Germany and one of the
founders of the European PTA collaboration. \\
Their talks gave us a broad overview of the past and current state of the PTA window on the GW universe. Caterina introduced pulsars and the effects of gravitational waves on the pulsar timing data. She discussed the most likely source of GWs that PTAs are sensitive to, the stochastic GW background produced by the incoherent superposition of GWs from several merging super massive binary black hole binaries (SMBHBs) and the various sources of noise in PTA experiments, such as intrinsic pulsar timing noise, Solar System ephemerides errors and noise from the interstellar medium and the Solar Wind. The talk ended by showing how the different PTAs are testing for the source of the common uncorrelated timing noise and some early hints that part of that signal might be the result of uncorrelated timing noise in individual pulsars. \\
This was followed by Michael who discussed the current state of the art in PTAs and their future. He touched on the common but spatially uncorrelated signal that is now emerging in the datasets of the three older PTAs (and the IPTA, the umbrella collaboration of all the PTAs) and clarified why this is not a detection of the GWB itself, yet. He then introduced the various upcoming radio astronomy facilities which will lead to a dramatic boost in the sensitivity of PTAs. This sensitivity will be crucial for PTAs to begin probing signals beyond the stochastic background, such as continuous GWs from single sources and the tests of gravity. He also talked about the various upcoming PTA experiments and the impacts emerging telescopes will have on the sensitivity regime.

\subsection*{Transcript of Q\&A -- relevant comments only}
\begin{enumerate}[leftmargin=0mm]
\item[Lijing] Thank you very much for the nice talks by Michael and Caterina. So now is the time for the discussion. So there are already many questions in the chat and some of them are already answered by Caterina or maybe does someone want to extend on these questions? Probably we have a left question that has not been answered yet, it's from Mikhail (Lijing reads) - "Could we send an observatory into space to mitigate some of the Earth and Solar system noise by separating the detectors and factoring the local noises out?"
\item[Michael] It doesn't help you because you still need to know the position of where your telescope is, whether it is on the earth or a spacecraft, you still need to know the relative position of the Solar System barycenter because that is the best inertial frame nearby so you still need that position it is not like moving LIGO into space to make LISA, that doesn't buy you anything, we still have the interstellar medium we still have the ephemerides problems, unfortunately not, so we can happily do this from Earth.
\item[B\'eatrice] Hi and thanks for these wonderful talks. Very informative and congrats on getting so close. I was wondering if you actually
observe now GWs from a single source how much can you tell about the source and its parameters? Can you tell the masses, spin, eccentricities or how much do you learn about the source itself?
\item[Michael] Yes, you can do exactly the same thing. You can get a chirp mass and so from that point of view you immediately see the orbital motion and then you will be able to tell something about the masses as well.
\item[B\'eatrice] So both masses?
\item[Michael] I think you will get the chirp mass more than anything else but then how to separate this I have to think about this but maybe Caterina knows the answer.
\item[Caterina] No I would have said the same, I think that you can get the chirp mass.
\item[B\'eatrice] \textelp{} and that's pretty much the only parameters you can get? No fun things like eccentricity and spins?
\item[Michael] Spins, no. No way. Eccentricity perhaps, depends. From the sort of cycle that you see, you'll see some higher harmonics but not purely sinusoid.
\item[Sathya] Can I just clarify? This is Sathya here. Hi Michael. Sorry that I could not join earlier due to earlier meetings. (Michael -- Good to see you!) To measure the chirp you need to have a change in frequency that is observed. So you are observing at nanohertz, how much change in frequency do you expect for your typical sources and that will determine how accurately you can measure (it)?
I think I was wrong, because you are saying you can measure chirp mass. I thought the frequency does not change at all, it appears like a continuous wave except that you have to worry about the rotation of the Earth and therefore a modulation of the signal but I wanted to understand how much change in frequency and are you talking about the highest frequencies also, for instance.
\item[Michael] Yes. I mean again you need long datasets, so yes, typically. That's why I put on my slide for most sources you only see a more or less constant frequency but if your dataset is long enough and your pulsars are oriented in the right way and you are not to far away in separation then you should be able to pick up some frequency change but it's really much source dependent, you will not be able to get this for all the sources, that's correct but as I said on my slide for some datasets for particular sources you may see an evolving frequency range. It depends of course on the sensitivity and on how good you can actually measure the signal. I mean everything is geared to the signal to noise of the detection.
\item[Sathya] The reason I mentioned is you see roughly a cycle every thirty years if you detect a new source you probably would need to wait for long enough so that is one constraint and in that sense I really like the approach of having longer cadence that you said was an experiment (I am sorry if I taking someone else's time) but do you have plans for that to be resurrected and take data at more cadence?
\item[Michael] I mean, I have to say we have not been good in coordinating that again because the telescopes are all overbooked, and so synchronizing the cadence is difficult. It is a practical thing, but if you have for instance a dedicated pulsar telescope or at least a few pulsar telescopes you can make sure they are interleaved, so that improves the cadence, so it's not impossible. To be honest when we did the experiment it wasn't on purpose. It was just after a while we realized "Oh wow, we get a really good cadence on this source" and that's when we looked at the dataset. So it is one of the standard brightest pulsar that you can look at so every telescope looked at it even if it was for calibration purposes but we could and should make this a more systematic approach but because of the over-subscription it was just not possible so far.
\item[Arianna] Sure, my connection is very poor so if I break out just read my question but I was really interested by the complementarity with LISA slide. I think this also came up before in the chat, I just wanted to understand better: you start off the with the source being in the PTA band, and then it goes into the LISA band but why is there a moment where you can also see it in the PTAs after? Is that some resonance or what's going on there?
\item[Michael] When you merge you basically get a step function in the space time change and that can be picked up as a phase jump in the rotation of the pulsar. So when you plot your residuals there will essentially be a phase step in your residuals and that's why momentarily when you see a merger event you should be able to see this. Time of arrivals (TOAs) do not line up any more but there's a step function and therefore because you continue to monitor for this source for the PTAs, there will be a moment where we will see this event happening. This particular moment is detectable in PTAs again.
\item[Caterina] \textelp{} and then it will basically modify space forever so it's really, yeah, well, a blink and you miss (it). Yeah, it is the memory effect (that was just typed in the chat).
\item[Lijing] Yeah I realized I had a delay in getting the messages in the chat. Luis, do you have a question?
\item[Luis] Hello, yes. Can you hear me? (Yes) So first of all thank you for these very nice lectures. I wonder, it was mentioned by Kramer, a way to get a better resolution is to get more pulsars. I wonder how reasonable it is to wait to get more pulsars to get the resolution we want or need. A more precise way to put my question; how many new pulsars do we detect per year and how many do we need to double the sensitivity of this experiment?
\item[Caterina] I was thinking to this study that was done a few years ago. If I remember correctly, the number of pulsar (used in the pulsar array) is the most affecting parameter in the sensitivity to a GW background. If I am not wrong, fifty to a hundred pulsars would be ideal. There is a study from 2013 that gives the various relationships, I can put that in the chat but concerning the number of pulsars that we can add to a PTA every year, I wouldn't know.
\item[Michael] Yes, I think one interesting thing about PTAs first of all to know is that in principle the sensitivity scales linearly with the number of pulsars and the easy way to explain it depends on the detection and that's the paper that Caterina mentioned. The rule of thumb is that because, indeed we have pairs of pulsars and the number of pairs goes as n-squared, and the sensitivity then goes as the square root of n-squared so you basically end with a linear improvement depending on the sensitivity range. Caterina pointed out the paper you need to read to understand all that but the number of pulsars we can add every day. So again the new telescopes will help enormously, because at the moment when I started my PhD we knew about, I don't know, three hundred or so maybe pulsars? Um, and now we know three thousand. But with new telescopes there's potentially about thirty thousand pulsars in the galaxy that we can detect. If you assume that 10\% of them are millisecond pulsars which is what we need, which is roughly the scale that you would find, so there's about three thousand (millisecond) pulsars to be detected. If we need about a hundred pulsars, as Caterina just said, there's a good chance that we get this number that are suitable not such a long distance (into the) future. So for instance we are doing a survey with the Meerkat telescope and we expect that we detect another 100 millisecond pulsars just with that survey and FAST is discovering many sources leading to papers like on a conveyor belt because of its sensitivity. So the number of discoveries has not slowed down, if anything it goes up, it increases and eventually you run out of Galaxy but the nice thing is with pulsar timing it doesn't actually matter, in principle, how far the pulsar is away as long as we use the Earth term. But eventually we want to use the pulsar term and the distance to the pulsar (then) matters but every millisecond pulsar has the potential to be absorbed in a pulsar timing array. So I'm quite optimistic that we will get to these hundred pulsars that we can put in quite soon, and then of course you need to have some timing baseline, but we are getting there.
\item[Lijing] Actually, if I may chime in, I am a little worried about the discovery of more pulsars, because if we have more pulsars, with a fixed telescope time we spend less time on each pulsar so how do you handle this? Do you have some strategy?
\item[Michael] Ah, we already do that. Of course with FAST is a waste almost to follow up all these pulsars discovered. That is why small telescopes like the 100-m dish we have helps FAST in timing those. So, give us some space Lijing, give us some work to do and we will try to follow up at least some of the pulsars. We can tell you then which ones FAST has to concentrate on for high precision timing but that's the nice thing about having many telescopes around the world. We can share the load but you are right, the more pulsars we discover, the more pulsars we have to follow up. But to be honestly brutal we already discover pulsars and as soon as we realize it is not interesting we stop timing them. It is already happening. So we really need to concentrate on the most promising ones.
\item[Golam] I just wanted to quickly, well, for the PTAs and pulsar people this is stuff you know, but there are two other factors which we often forget. One is that pulsar discoveries are heavily reliant on computing power and if computing power scales, then you suddenly might just take old data and rediscover a whole bunch of pulsars, which Michael can tell you about I think: he's been on a fair number of these papers where you just take old survey data and you keep finding new pulsars. Then there is this purely statistical phenomenon that we keep finding amazingly bright, nearby extremely good PTA sources, every so many months and it's hard to figure out why we haven't seen them before but we keep finding more and more of these pulsars and, I think Michael showed on his slides that you need about five years of timing data on some of these. Eventually it is more likely to be a question of the problem that Lijing brought up, which is that how do you find the telescope time to track these sources rather than worry of not having enough sources on the sky?
\item[Michael] We need more help! We need more man/woman/human power, whatever, to help us.
\item[Garvin] Yes, alright okay, yeah thank you. Thank you for the very interesting talks. I am very interested in the intrinsic spins of pulsars and you mentioned before that the intrinsic spin may cause some of the residuals that we see. I was just wondering if that is the same as jitter noise that I have read about, and if so are there models for this? What are the leading models?
\item[Caterina] Yeah, sure. So, we know that jitter noise is going to be a limiting factor. This is going to be a problem in the moment in which we have more sensitive telescopes. There have been some studies looking into this in particular, although I think right now we are basically taking care of jitter noise with specific parameters in the pulsar ephemeris but there have been attempts to develop techniques specifically dedicated to mitigate the impact of jitter. In particular there has been a study in 2014, if I remember correctly, based on single pulses. So instead of using an entire observation to determine one TOA you work on the stream of pulses that you are receiving during the time of the pointing. Then, this study applies principal component analysis that, if also used on the full Stokes observations, not only on the total intensity, is able to reduce the impact of jitter noise if I remember correctly, by 40\%. I don't think that it is currently applied (regularly in PTAs) but there is this possibility. Of course, the problem of this technique is that it relies on single pulses and that the amount of data that you retain with respect to what you usually do is enormous, so it is a problem. You have to find a bit of a trade off. Many of the techniques would be wonderful if we had, like, infinite disk space, of course. So there is the possibility (of mitigating for jitter noise) but I don't think it is implemented by anyone, as far as I know. Michael, I am talking of Stefan's technique.
\item[Garvin] So the way you talked about it then, it sounds almost like jitter noise is defined as a mixture of instrumental and actual magnetospheric effects. Is there no distinction within jitter noise the actual intrinsic effects and possible instrumentation jitter?
\item[Michael] The instrumentation is not a problem. We can measure the arrival times to nanosecond precision. It's really intrinsic, because the pulse width is finite and the emission is random from pulse to pulse so it jitters within your window. That's the jitter noise. On top of that you have actually other more, other red noise processes intrinsic to the pulsar which comes either from the magnetosphere or from the interior of the neutron star, both of which we are not 100\% clear. Certainly we see the magnetosphere having an impact on some young pulsars but for the old, millisecond pulsars it is much more difficult to tell. But instrumentation is clearly not the problem. As Caterina says, keeping the data, the baseband data that you probably want is a technical problem but it's not an instrumental issue as such. It is the logistics, was what I wanted to say. It is not a technical problem.
\item[Lijing] So actually I have a theory about noises. These are noises because we don't understand them. If we have an understanding of them they can be signals. Actually, there is a famous paper by Andrew Lyne et al. from 2010. They tried to understand the noises. So what's the prospect in understanding these noises as signals or something like this?
\item[Michael] Yes, so what we did in that paper we showed that some distinct changes in the magnetosphere cause a change in the spin down rate and if you have enough cadence (back to cadence) to actually determine at which point it switched from one spin down rate to another spin down rate, and you will even be able to remove the timing noise completely. We may even be able to use normal pulsars rather than millisecond pulsars for PTA purposes. Yeah, so we proposed that but honestly we just didn't have the cadence to actually give that a try, but in principle it's possible
\item[Sathya] Yeah thank you I want to further follow up whether it is noise or something else which of the different noise sources that you discussed scale in the same way with time, because depending on where these sources are probably they have different dependence on time and in particular I am thinking, over a thirty year period when you have the data, wouldn't, lets say, our direction to pulsars change such that you may have a different dependence on time of the interstellar medium? I know it doesn't change a lot but, you know, at the same time, radio telescopes have phenomenal resolution like milli or micro arc-seconds. So I was wondering about those questions but I have one more comment to make after that.
\item[Caterina] How long you should observe before distinguishing the various sources of noise, it depends basically on the shape of the power spectrum, in the sense that of course, I think the GWB is the steepest noise source that we have (in terms of the theoretical function form). For example the noise induced by the interstellar medium is not as steep but the amplitude is higher. So it would cover the signal from the GW (for timing noise I am \textelp{} like intrinsic timing noise, I am not an expert so I am not really sure what is the expectation for the slope of the power spectrum). But I think that the problem with the other sources of noise like the correlated ones, such as the solar system ephemeris, the issue that we found is that, as I was showing before, both the amplitude and the slope of the spectrum are quite similar to the gravitational wave background (ones). So, I don't think that it is a solution to say "wait to see which one emerges" because in the end some of them are so competitive in terms of characteristics in the power spectra that it would not be super useful, I think.
\item[Sathya] Knowing the position of Jupiter and your own telescopes, will future technology or, I don't know, ephemeris data with new satellites and so on, would that help?
\item[Michael] Yeah it would definitely help. I mean I didn't mention it, the reason why, I mean I pointed out that the upper limits people published in the past are way below the kind of common red signals that is now seen by PTAS, the reason why that could the case is because now we have covered enough of the orbital phase of Jupiter for the signal to decorrelate while before that it may have been absorbed in some ephemerides not covering the full Jupiter orbit and with every ephemerides set that is being published we will improve our knowledge about the solar system ephemerides and there is some nice work being done in PKU and Kavli by KJ and Yanjun Gao. They basically try to solve for the planetary ephemerides and the planets at the same time. And that's ultimately, I think, the way to do it. I mean the other approaches like Bayesephem and so on, you try to (I mean, Caterina knows this better than me) but the whole idea I think at the end is that you need to solve the pulsars and ephemerides all in one go.
\item[Sathya] Now, the final comment I had was I thought the very purpose of things like SKA and MeerkAT is to observe the whole sky all the time and therefore having more pulsars in principle is not a problem. It is a data analysis problem but if you have good reasons why something can be followed up, the cadence can improve and you can accommodate as many pulsars as you want. Is that not right? Am I getting something wrong there?
\item[Michael] No, it is absolutely right. You should give us the telescope time. If you are in charge that would be great. I am not sure you can convince the other communities but I mean \textelp{}
\item[Lijing] So Michael and Caterina, I have a tough question for you two. We know that in 2015 LIGO detected the first event of binary black hole merger and the announcement was in 2016. But that year the Nobel prize didn't go to the discovery and it was only one year after because there was a detection by LIGO and VIRGO, the GW event GW170814. Then the Nobel prize goes to just three people, three leading people in the collaboration. For the PTAs will you encounter similar situations? You know you don't have really an independent cross-check probably or do you think the PTAs in themselves can provide some cross check?
\item[Caterina] So, currently, yes we are taking this quite cautiously in the sense that of course, as you say, PTAs are quite of a \textelp{} a very delicate experiment, we don't have control on our detectors. So what we are doing, in the EPTA at least, I am not sure in other collaborations, is to make sure to use different software to compare whether they are giving the same answer or not. We are checking whether the signals that we are seeing for example evolve in time. So, currently we are working on a second data release for the EPTA and we are making sure that the results that we are getting come as a natural evolution of what we saw in the first data release. So of course, these are all internal checks, but at least the usage of different and completely independent software, and different detection approaches, builds some confidence in what we see. However we aim also, in the next steps, to have some external input from the GW community and build some more confidence.
\item[Michael] I think Sathya wants to point out that there was no correlation to the VIRGO participation. I think Lijing the detection was too late, the nominations had already gone in at the time when that detection was made. Um, but, yeah, Caterina's absolutely right. I think we do need cross checks. However, personally I don't think it will be a Nobel prize. I think gravitational wave detection has been done twice now already, with the Hulse-Taylor pulsar and with the direct detection. So I don't think it's adding anything to the race. We lost the race. There was a time when maybe PTAs had a fighting chance to be there before LIGO but I think that's gone. That's fine, we can concentrate on the science, that's very good. I think the main thing is of questions that Caterina already says, we need the independent cross-correlation (which) at the end of the day we don't have. We observe the same pulsars, so the best thing we can do is to observe with different telescopes and compare, and of course, (with) the IPTA putting everything together, everything should become more significant. So that is another test somehow but within Europe, I think I indicated in my talk, we have learned a lot by comparing our telescopes and PPTA and NANOGRAV, because Arecibo and NANOGRAV (GBT) did observe different samples of sources, only one or two pulsars overlap so a lot of things can happen. While in the EPTA, well we have more data to analyze and that's more work and so on but it was helpful because we saw all the dirt. All the instrumental effects that can happen, and there's a lot! And it is easily swept under the carpet or not noticed, let's put it that way, if you have only one telescope. If you have a variation in your clock, it may be very subtle and you may not notice but if you have different telescopes and you observed with them simultaneously once a month with LEAP where we have to phase them up to a small fraction of a second, um, you notice things and that is super helpful and I think that level of comparison we need to achieve with the IPTA to be sure that what we see is real. I think we need to demand this from us, and the community should demand this from us. They should not be believing anything until we have shown this. I think that is my firm, firm belief.
\item[Golam] I'll sort of quickly chime with my IPTA hat and point out that we do try a lot of statistical testing to show that we are not unusually biased towards any particular signal, so for instance we have the dropout method where essentially we pretend that instead 50 pulsars in the IPTA we might have 20 or 30 or 40 pulsars and then try to look at whether we would recover the signal, albeit with a lower amplitude or sensitivity from those many different combinations but as we are getting closer to the critical junction we are also looking into our analyses, well our GWA pipelines as well and now we are looking very seriously at some of the effects of things which we haven't had to worry about for a long time. For instance our sampling routines or many of our Bayesian approaches we've previously agreed on as 'this is possible, this is what works. lets just go with this.' and now we are going back and digging deep into things we as a large body of pulsar astronomers into the statistical end of things and one of the comments is actually very relevant to some of the kind of new things we are trying. Which is the question about the slope, of the GWB. So the $-2/3$ value, it's just a canonical value but now we are starting wonder about what happens if you let it wander around a little bit but there is a lot of input that's coming in from the GW community as well, we are getting closer. I for instance am sitting in a group which is almost entirely GW and I am just a pulsar astronomer; so this is an ongoing process. I'm sorry I dropped out of the call so I might have taken you off on a bit of a tangent but yeah, we do worry about a lot of things, especially at the IPTA and we are trying a whole bunch of different things, within the EPTA as well ad every PTA is equally skeptical of it's own results.
\item[Lijing] Thank you very much for the answers. I am very happy to hear that the cross-checks are well performed and it is guaranteed that the result is very, very solid. So a related question, does this measurement, if we have measured it, can be cross-checked with other electromagnetic, uh, other observations like from optical or others? For example can we check if the underlying population is consistent from optical observations and the GW observations, from PTAs. Is it possible?
\item[Michael] Caterina do you want to? Well, in principle, yeah, there have been (attempts) lets put it the other way around people systematically search for optically identified binary BH systems, and then see if they can see them in our dataset. that would be the easiest way of doing it if you know what the orbital period is you can try to find it in your data. That has not been successful. Chiara Mingharelli has done some very nice work in this line working very closely with electromagnetic people, trying to identify. Alberto Sesana has some very nice simulations about how often the signals would be in our datasets and so on, and the simulations may be calibrated by optical or X-ray or other observations can give you a feel for the level of the signal from the stochastic background. But ultimately I think it would be nice to have a single source detection to really do what you suggested to do. Eventually, I think if we don't find a single source with enough sensitivity then I think there will be a question asked. I think the fact is by the way as we pushed our limits lower and lower theoreticians of course adopted their models for merging galaxies and the astrophysics with it, and eventually some of them must be wrong and some of them must be right. This is the nice feedback. Golam mentioned the spectrum: there is so much astrophysics possibly modifying the spectrum that by actually putting limits, frequency dependent limits which we don't usually quote, we usually put it at 1/year, but if we put actually frequency dependent limits you can actually narrow down the astrophysics that is governing galaxy merger. It is a nice little feedback loop between, I think, different optical observations, X-ray, theoreticians, pulsar timing people. I think we are only seeing the beginning. I think once our data are much better and we have much better frequency coverage of our spectrum I think lots of the nice things you mentioned can be done.
\item[Nicola] I am already surprised by the fact that you can actually detect the memory and this actually a source of the kind of multiband analysis that you can make. I didn't know this and it is quite interesting to know that you can do this with LISA and PTA together.
\item[Michael] I think if I am not mistaken the first time this was pointed out was by Yuri Levin with Rutger van Haasteren in Leiden university when Yuri was still there and I remember that was the first time I ever heard this and when they mentioned it, I did not understand it but yeah you can measure this in your residuals.
\item[Nicola] Is it something you can do because you have a multiband detection, in the sense that maybe you see the merger in LISA and based on this observation you can look in the PTA data and see if there is a signal also there?
\item[Michael] You should see this anyway in the data, if you have a milli-second pulsar which has a step change and maybe you see another pulsar in a certain geometry or direction then you should still be able to, independent of any LISA detection, you should still be be able to see the step change. People have looked at this and found nothing so far. There is a nice paper by NANOGRAv who looked at their dataset quite recently to have a systematic search for the burst with memory effect and that is still an upper limit.
\item[Nicola] Since I am showing my face and I have the floor here: you showed a slide Michael about possible environmental effects on the slope of the stochastic background that can change based on different environmental effects like eccentricity and stellar hardening and so on, are these effects all suppressing? From your slide it seems that these effects all suppress the background. Is there any chance that they can either increase it or give it different bumps or feature that we can look for in the background data?
\item[Michael] I'm not sure if Caterina has a better answer than that but in most cases they tend to make things more difficult than help. I think Alberto Sesana was the first to point this out and I remember this very vividly: I was visiting the Australian colleagues and this paper came out "oh sh..., our signal is going to disappear!" and it was uh, interesting. The flip side that you can learn something about the astrophysics by doing PTA experiments but it brought the thinking about which frequency range we have to look back to our minds.
\item[Lorenzo] I was wondering if in connection to this question we can actually once we will have maybe the detection, make constraints on these aspects from the data so if we can actually understand all this physics beyond the production of GWs.
\item[Michael] Yeah, I think that is exactly what is being tried. The problem is (to my naive thinking, maybe Caterina or someone else can correct me) that we don't know quite the absolute level of the spectrum. If we were to know the absolute level of the spectrum then you can, by having frequency dependent limits, pinpoint the astrophysics a bit more accurately. But that is my naive thinking maybe someone else can improve on that answer but the basic idea is yes you can infer the astrophysics from those observations. In fact if people wouldn't have thought about the limits coming from PTAs I think it would have been difficult for them to realize that the spectrum is actually changing. I think that was motivated by the fact that PTAs have not seen anything, so again there is this nice feedback between that and the theoreticians.
\item[Lijing] Actually, I have a question. In Michael's talk about the LEAP project, you mentioned by combining five telescopes in Europe to form an effective larger telescope. So what is the advantage of having this effective larger telescope instead of five telescopes? If you have five you have more, but less sensitive.
\item[Michael] Sensitivity of course is the short answer. Um, finding out systematics, for instance, in the timing at each telescope, phasing them up. If you observe the same source simultaneously with five different telescopes and you don't see the same thing, something is wrong. It's very helpful to identify instrumental effects. Also these nice observations of the interstellar medium, the scattering measurements where we basically actually image the interstellar medium. You can basically take an image of the interstellar medium at the same time as you time the pulsar, that's another advantage of combining these five telescopes in a coherent array. On top of that, we can still analyse the data recorded at a single telescope during the LEAP observations independently, so then you can also have the single telescope dataset compared with the joint dataset so you can add sensitivity and cross-checks and the interstellar medium  which makes it such a nice tool. We actually tried to combine FAST with LEAP and we have taken some data and we are still analyzing it, but combining FAST and LEAP we get almost a 400-m dish. FAST illuminated is 300 but with LEAP added it could be something like a 400-m dish and yeah, so that's the biggest pulsar telescope in the world. If you combine it with more FAST telescopes, then it limits (us) to the rotation of the Earth how long we can see a source with all the telescopes at the same time. LEAP is a very nice tool. It has been quite helpful.
\item[Lijing] So what is the most difficult technological part of combining these telescopes?
\item[Michael] Technically it is not difficult. People do VLBI all the time but what we actually had to do is that we had to write our own correlator, we had to write our own calibration. The thing is, and here we go down to radio astronomical techniques, each telescope usually has a receiver and one has linear feeds, one has circular feed and the biggest challenge is good calibration. As (in) almost every experiment good calibration is the challenge. You don't reach 100\% coherency if you don't calibrate the telescope before you add them. The way we did that I think that it is quite a novel idea, VLBI does not do it. We calibrate each telescope on the baseband data before we add them and so for instance (hopefully the EVN don't kill me) but EVN has tried to do the same experiment that we do with LEAP more or less as a byproduct and because they do the calibration differently they have never been actually been able to this but with LEAP we do this routinely.
\item[Golam] So we have about 8 minutes left and I wanted to shift the focus a little bit from science to some of your personal experiences or your insights and opinions on how you see PTAs shaping up in the GW, well, in a sense, a spectrum; this convergence of fields that is birthing, rather, has birthed a new field.
\item[Golam] Well, it's a sort of loosely worded question but I was wondering what you think of where PTAs are headed in the GW landscape and particularly if there are early career scientists interested in entering the field, what are your experiences or what are the words of wisdom you would pass on to them?
\item[Caterina] Well OK, I mean, PTAs are the only experiment that can open the nanohertz window (for the time being) and I think that of course then the challenge, but also the beauty of GW astronomy, will be to coordinate the results from all they experiments searching for GWs, when we will all achieve a detection. That is going to be massive. On the PTA side (I am not a theorist, so I am not sure of what is the next evolution in the theory field) and concerning the data analysis and data analysis techniques, I think that there is a humongous amount of work to do, so there a lot of opportunities (to enter the PTA field). For example from the point of view of low frequency pulsar astronomy and interstellar medium and solar wind mitigation, that is the field where I am currently working the most.
\item[Golam] In the last couple of minutes I will quickly go to Michael (and maybe Lijing you can round this off) with any insights of how it is to be remotely collaborating with people literally in every part of the world and doing distributed data analysis, and trying to figure out who did what and how it worked, all together.
\item[Michael] First of all, I think these are strange times. I would rather be and visit people and sit with them at the same table or look at the same screen to do things. That's much more fun, and better but I think everyone will already agree to that. I think it is, well, to be honest the reason why I continued with astronomy after my PhD was because I met so many wonderful people. The science was great but the people were even better, and that was my motivation. Being in a collaboration as big as the IPTA you always have people in the collaboration you don't like, that much like in a big family: some family members you don't really like that much but there are lots of people you really like and want to spend time and work with, and that I think is the greatest motivation for me. Science is great but actually having the chance to meet old people, young people, new people every time I think that keeps us going and just do pulsars! With pulsars you can do so much, you can study gravity, you can study the interstellar medium, the galaxy and you discover new things. It is just nice, so I think that combination of it having a rich field where you can just do so many things with pulsars where you meet bright young and old people from all backgrounds and all cultures that is what a collaboration is really stimulating for me and I wouldn't (want to) miss that.
\end{enumerate}

\subsection*{Conclusion}
PTAs are poised on the brink of some of the most exciting results since pulsar astronomers started searching for the tell-tale signature of GWs on pulsar signals. There is a very strong hope that the detection of the stochastic GW background (GWB) is imminent. Although it is challenging to observe single sources with the current state-of-the-art, new radio astronomy instruments such as the Square Kilometer Array (SKA) are expected to truly open the window on the nano-hertz GW regime, offering not just a unique probe of the most massive black-hole binaries but also true complementarity in the micro-hertz regime for land and space-based interferometric GW instruments. \\
The talks gave us a broad overview of the scientific possibilities and the lively discussion delineated the possibilities and limitations of PTAs as instruments. However, the conversation also reminded us that ultimately, the pursuit of GW science is a human endeavour and its greatest capital the spirit of enquiry. 

%\printbibliography
\end{refsection}

\end{document}